\documentclass[12pt,english]{article}
\usepackage{amssymb}
\usepackage{amsfonts}
\usepackage{amsmath}
\usepackage[authoryear]{natbib}
\usepackage[T1]{fontenc}
\usepackage[latin9]{inputenc}
\usepackage[letterpaper]{geometry}
\usepackage{amsthm}
\usepackage{setspace}
\usepackage{babel}
\usepackage[normalem]{ulem}
\usepackage[hidelinks]{hyperref}
\usepackage{version}
\usepackage{graphicx}
\usepackage{url}
\usepackage{color}
\usepackage{multirow}
\usepackage{float}

\numberwithin{equation}{section}
\newtheorem{assumption}{Assumption}[section]

\newtheorem{proposition}{Proposition}[section]
\newtheorem*{note*}{\protect\notename}
\renewcommand{\cite}{\citeasnoun}
\allowdisplaybreaks

\makeatletter
\theoremstyle{plain}
\newtheorem*{thm*}{\protect\theoremname}
\theoremstyle{remark}
\newtheorem{rem}{\protect\remarkname}
\theoremstyle{plain}
\newtheorem{lem}{\protect\lemmaname}
\theoremstyle{plain}
\newtheorem*{lem*}{\protect\lemmaname}
\theoremstyle{remark}
\newtheorem*{rem*}{\protect\remarkname}
\theoremstyle{plain}
\newtheorem{thm}{\protect\theoremname}
\makeatother

\providecommand{\lemmaname}{Lemma}
\providecommand{\remarkname}{Remark}
\providecommand{\theoremname}{Theorem}
\providecommand{\theoremname}{Theorem}
\providecommand{\notename}{Note}
\hypersetup{
	linkcolor=magenta, urlcolor=blue, citecolor=blue, unicode=true, colorlinks=true}

\setcitestyle{authoryear,open={(},close={)}}
\begin{document}
	\pagenumbering{arabic}
	\title{Predictive Quantile Regression with High-Dimensional Predictors: The Variable Screening Approach}
	\author{Hongqi Chen\thanks{
			College of Finance and Statistics, Hunan University. Email:
			hongqichen@hnu.edu.cn } \and Ji Hyung Lee \thanks{
			Department of Economics, University of Illinois Urbana-Champaign. Email:
			jihyung@illinois.edu} }
	
	\maketitle
	
	\begin{abstract}
		This paper advances a variable screening approach to enhance conditional quantile forecasts using high-dimensional predictors. We have refined and augmented the quantile partial correlation (QPC)-based variable screening proposed by \citet{ma2017variable} to accommodate $\beta$-mixing time-series data. Our approach is inclusive of i.i.d scenarios but introduces new convergence bounds for time-series contexts, suggesting the performance of QPC-based screening is influenced by the degree of time-series dependence. Through Monte Carlo simulations, we validate the effectiveness of QPC under weak dependence. Our empirical assessment of variable selection for growth-at-risk (GaR) forecasting underscores the method's advantages, revealing that specific labor market determinants play a pivotal role in forecasting GaR. While prior empirical research has predominantly considered a limited set of predictors, we employ the comprehensive Fred-QD dataset, retaining a richer breadth of information for GaR forecasts.
		\newline
		\emph{Keywords: variable screening; high dimensional time series; $\beta$-mixing; quantile regression; growth-at-risk.}
		
	\end{abstract}
	\footnote{This paper is the part of Hongqi Chen's Ph.D dissertation at University of Illinois Urbana-Champaign. We are deeply indebted to the dissertation committee members, Marcelo Cunha Medeiros,  Xiaofeng Shao and EunYi Chung for their support and guidance. We are also grateful to  seminar participants at UIUC and AMES 2023.}
	\newpage
	
	\section{Introduction}
	
	Stability in economic growth is gaining prominence in macroeconomic research and policy discussions. To quantify tail risks and forecast recession levels, researchers, exemplified by \citet{adrian2019vulnerable}, have introduced growth-at-risk (GaR) as a metric, defining the (conditional) quantile of GDP growth rate. 
	
	Forecasting conditional quantiles of economic time series is complex due to macroeconomic intricacies and abundant data. There's ongoing debate regarding the predictability of GaR. For instance, \citet{adrian2019vulnerable} employ the National Financial Conditions Index (NFCI) with quantile regression to forecast US GaR. They find this index effective for predicting potential recession levels. Subsequently, quantile regression using NFCI gained popularity in GaR research. \citet{adrian2020term} apply panel quantile regressions to analyze 11 advanced economies. Additionally, they are constructing a term structure for GaR that captures the intertemporal tradeoff. \citet{figueres2020vulnerable} study vulnerable growth phenomena in the European area. Beyond NFCI usage, \citet{brownlees2021backtesting} demonstrate that the GARCH model outperforms quantile regression in GaR forecasting. \citet{plagborg2020growth} argue that the robustness of GaR estimation can be questioned when endogenous variables, notably NFCI, are utilized as extracted factors. These debates highlight the need for careful variable and model selection when predicting GaR and economic time series distributions.
	
	Considering the importance of variable selection, a fitting approach for high-dimensional datasets is essential, particularly in modern econometrics and statistics research during the era of big data. Forecasters often face with an abundance of potential predictors, including hundreds or even thousands of macroeconomic and financial time series. It is therefore crucial to identify the most important predictors through variable screening or selection methods. In this paper, we present our screening approach tailored for quantile regressions with time series data. There are three advantages of the variable screening approach. First, it excels in handling ultra-high dimensional data. Second, it offers computational efficiency, particularly with quantile regressions. Additionally, unlike $l_1$-penalization methods, it does not require stringent conditions for variable selection consistency. In the comparative simulation analysis, our variable screening approach outperforms penalization methods.
	
	Variable screening efficiently reduces the dimensionality of high (or ultra-high) dimensional datasets. In contrast to penalization methods, variable screening typically employs a two-step process. Initially, it screens the variables to narrow down the selection set, and subsequently, it selects the most significant predictors using information criteria or penalization methods. This leverages the well-known screening consistency property, as demonstrated by \citet{fan2008sure}, enabling the identification of the true set of predictors. This screening consistency proves advantageous in the context of the two-step process.
	
	Extensive literature exists on variable screening. Among the earliest methods is sure independence screening (SIS), introduced by \citet{fan2008sure}, and \citet{wang2009forward} demonstrates that forward selection can serve as another viable option for variable screening. While numerous studies concentrate on least squares regression, \citet{ma2017variable}  introduces three screening algorithms relying on quantile partial correlation (QPC), initially conceptualized by \citet{li2015quantile} as a quantile correlation-type measure. Furthermore, researchers like \citet{kong2019screening} and \citet{zhang2018variable} have introduced two more variants of quantile correlations tailored to accommodate specific model structures. In addition to quantile-type correlations, alternative correlation structures have been utilized for variable screening. To name a few, examples include the use of distance correlation by \citet{li2012feature} and martingale difference correlation by \citet{shao2014martingale}. These alternative approaches expand the toolkit available for variable screening, offering researchers a range of methods to suit their specific modeling needs.
	
	While the majority of variable screening methods for mean or quantile regressions are designed for i.i.d. data, few have extended these techniques for other structures, like dependent data. For example, \citet{yousuf2018variable}, the only paper on this topic, extends Sure Independence Screening (SIS) to high-dimensional time series models using the functional dependence framework by \citet{wu2005nonlinear}. We also mention other variable selection methods for dependent data, as our screening approach shares similarities with forward selection-type algorithms. Notably, \citet{ing2020model} and \citet{chiou2020variable} introduce orthogonal greedy algorithms for linear regression with dependent observations, while \citet{sancetta2016greedy} explores greedy algorithms for selecting relevant covariates under mixing conditions.

	In this study, we extend the QPC approach originally introduced by \citet{ma2017variable} to accommodate $\beta$-mixing time series data. To the best of our knowledge, prior to this study, there has been no research on variable screening in quantile regression involving dependent data. Our theoretical contributions encompass two key aspects. Firstly, we establish a refined convergence bounds for QPC in the context of i.i.d. data. Secondly, we derive new bounds for QPC under the stationary $\beta$-mixing condition and illustrate that the convergence of the screening method is influenced by the level of dependence (mixing coefficient). These findings indicate that higher dependence induces slower convergence, with the i.i.d. scenario included as a special case. Furthermore, the variable screening property is shown to remain valid. Expanding the QPC method from independent to dependent processes broadens its applicability for various time series applications. We employ the QPC approach for GaR prediction, utilizing the Fred-QD datasets introduced by \citet{mccracken2020fred}. Overall, our findings contribute to the advancement of more robust and effective methods for variable screening in the context of time series quantile regression.

	The paper is structured as follows: Section 2 introduces the variable screening in quantile regression. Section 3 presents the theoretical results under stationary $\beta$-mixing framework. In Section 4, we provide simulation studies confirming QPC's screening performance for time series data. Section 5 presents variable selection results for GaR forecasting with high-dimensional Fred-QD datasets. The last section concludes.

	We employ the following notations throughout the paper. We denote $%
	\rho_{\tau}\left(u\right)=u\left(\tau-1\left(u<0\right)\right)$ as the
	quantile loss function, and let $\psi_{\tau}\left(u\right)=\tau-1\left(u<0%
	\right)$.  $\lambda_{\max}\left(A\right)$ and $\lambda_{\min}\left(A%
	\right)$ denote the largest and smallest eigenvalues of matrix $A$. $\left\Vert
	a\right\Vert $ refers to the $l_{2}$-norm of vector $a$ and $\left\Vert
	A\right\Vert $ is the Frobenius norm of matrix $A$. We use $c$, $c_i$, $C$, $C_i$, $C^{*}$, and $C^{**}$
	with $i=1,2,\dots$ to denote generic nonnegative constants across the paper unless otherwise specified.
	
	\section{Model Framework and QPC Screening Approach}
	
	We consider the sequence of random variables $\{Y_{i+1},X_i\}_{i=1}^{n}$, where $\left\{
	Y_{i+1} \in \mathbb{R} \right\} _{i=1}^{n}$ is the response variable sequence, and $\left\{
	X_{i}\right\} _{i=1}^{n}$ is a stationary $\beta$-mixing sequence of $p$-dimensional covariates. We represent each $X_i$ as $X_{i}=\left( X_{i,1},\dots ,X_{i,p}\right) ^{^{T}}\in \mathbb{R}^p$ with $p>n$. We aim to select covariates to estimate the conditional quantile of $Y_{i+1}$ given $X_i$ with a linear structure:
	$$Q_{Y_{i+1}}\left( \tau |X_{i}\right) =X_{i}^{T}\beta _{\tau }$$
	for any $i=1,\dots ,n$. We exclude the $\tau$ subscript in $\beta _\tau$ below for simplicity. To prevent notation ambiguity, we use $j$ instead of $i$ on $X_j$ to represent a particular covariate $\left( X_{1,j},\dots ,X_{n,j}\right) ^{^{T}} \in \mathbb{R}^n$ for $j=1,\dots ,p$.
	We use $S\subseteq \{1,\dots ,p\}$ to denote a generic
	index set of the covariate $X_i$. Given a index set $S$, $X_{i,S}$ indicates the covariates with the index set $S$ on $i$-th observation.
	Using variable screening to identify important covariates. We refine the QPC approach in \citet{ma2017variable} with weakly dependent data. Similar to the existing literature, variable screening depends on the forward stepwise selection technique, where one relevant covariate is chosen in each step. Our paper adopts two algorithms from \citet{ma2017variable} and details can be found at the end of this section. As these algorithms work step by step and rely on the QPC of each covariate, we can consider selecting covariate $X_{j}$ with the highest QPC, given the previously selected index set $S_{j}$, at a particular step $j$. In this case, we define the quantile partial correlation at $\tau$-th quantile level for random variables
	$\left\{ Y_{i+1},X_{i,j},X_{i,S_{j}}\right\} $ in step $j$ as: 
	\begin{align*}
		qpcor_{\tau}\left(Y_{i+1},X_{i,j}|X_{i,S_{j}}\right) & =\frac{cov\left(\psi_{\tau}\left(Y_{i+1}-X_{i,S_{j}}^{T}\alpha_{j}^{0}\right),X_{i,j}-X_{i,S_{j}}^{T}\theta_{j}^{0}\right)}{\sqrt{var\left(\psi_{\tau}\left(Y_{i+1}-X_{i,S_{j}}^{T}\alpha_{j}^{0}\right)var\left(X_{i,j}-X_{i,S_{j}}^{T}\theta_{j}^{0}\right)\right)}}\\
		& =\frac{E\left[\psi_{\tau}\left(Y_{i+1}-X_{i,S_{j}}^{T}\alpha_{j}^{0}\right)\left(X_{i,j}-X_{i,S_{j}}^{T}\theta_{j}^{0}\right)\right]}{\sqrt{\tau\left(1-\tau\right)\sigma_{ij}^{2}}}
	\end{align*}
	where $\alpha _{j}^{0}=\arg \min_{\alpha _{j}}E\left( \rho _{\tau }\left(
	Y_{i+1}-X_{i,S_{j}}^{^{T }}\alpha _{j}\right) \right) $ denotes the quantile regression coefficient of $X_{i,S_{j}}$ on $Y_{i+1}$, $\theta _{j}^{0}=\arg
	\min_{\theta _{j}}E\left( \left( X_{i,j}-X_{i,S_{j}}^{^{T }}\theta _{j}\right)
	^{2}\right) $ denotes the OLS coefficient of  $X_{i,S_{j}}$ on $X_{i,j}$, and $\sigma _{ij}^{2}=var\left( X_{i,j}-X_{i,S_{j}}^{^{T }}\theta
	_{j}\right) $ is the variance of the OLS residual of $X_{i,S_{j}}$ on $X_{i,j}$.
	
	The advantages of QPC are noteworthy. Firstly, QPC has a close relationship with population quantile regression coefficients.  Given our assumption of the linear conditional quantile form of $Y_{i+1}$ on $X_i$, we can define these coefficients as the minimizers of the quantile loss: \begin{equation*}
		\left(\beta _{1}^{0},\dots,\beta _{j}^{0},\dots,\beta _{p}^{0} \right) = \underset{\left(\beta _{1},\dots,\beta _{j},\dots,\beta _{p} \right)}{\arg \min }E\left( \rho _{\tau }\left(
		Y_{i+1}-\beta _{1}X_{i,1}-\cdots -\beta_{j }X_{i,j}-\cdots -\beta _{p }X_{i,p}\right) \right) 
	\end{equation*}
	
	In algorithms, given $\alpha _{j}^{0}$ and $S_j$ at step $j$, we can define $\beta _{j}^{\ast }$  as the following: \begin{equation*}
		\beta _{j}^{\ast }=\underset{\beta _{j}}{\arg \min }E\left( \rho _{\tau
		}\left( Y_{i+1}-X_{i,S_{j}}^{T}\alpha _{j}^{0}-X_{i,j}\beta _{j}\right) \right) .
	\end{equation*}
	
	According to \citet{li2015quantile}, we can express $qpcor_{\tau}\left(Y_{i+1},X_{i,j}|X_{i,S_{j}}\right) =\rho \left( \beta _{j}^{\ast }\right) $, where $%
	\rho \left( \cdot \right) $ is a continuous increasing function, and $%
	\rho \left( \beta _{j}^{\ast }\right) =0$ if and only if $\beta _{j}^{\ast
	}=0$. Furthermore, the lemma below establishes that $%
	\beta _{j}^{\ast }=0$ if and only if $\beta _{j}^{0}=0$. Consequently, it is straightforward to use $qpcor_{\tau}\left(Y_{i+1},X_{i,j}|X_{i,S_{j}}\right) $ as a substitute for $\beta _{j}^{0}$ to rank the importance of covariates. This valuable connection holds regardless of the data dependency structure.

	\begin{lem*} (Lemma 1 in \citet{ma2017variable})
		\label{lem:1} Suppose $\beta^{0}=\left(\beta_{0}^{0},\dots,\beta_{p}^{0}%
		\right)^{T}$ is the unique minimizer of $E\left(\rho_{\tau}\left(Y_{i+1}-%
		\beta_{0}-\beta_{1}X_{i,1}-\cdots-\beta_{p}X_{i,p}\right)\right)$. Suppose $%
		\alpha_{j}^{0}$ are unique minimizers of $E\left(\rho_{\tau}%
		\left(Y_{i+1}-X_{i,-j}^{T}\alpha\right)\right)$ and $\left(\beta_{0}^{*},%
		\beta_{j}^{*}\right)^{T}$are unique minimizers of $E\left(\rho_{\tau}%
		\left(Y_{i+1}^{*}-\beta_{0}-X_{i,j}\beta_{j\tau}\right)\right)$ where $%
		Y_{i+1}^{*}=Y_{i+1}-X_{i,-j}^{T}\alpha_{j}^{0}$. Then $\beta_{j}^{0}=0$ if
		and only if $\beta_{j}^{*}=0$.
	\end{lem*}
	
	Secondly, the property of the QPC approach becomes particularly attractive when $p>n$. Directly estimating $\beta _{j}^{0}$ when $p>n$ is unfeasible without introducing parameter penalizations. However, we can consistently obtain $qpcor_{\tau}\left(Y_{i+1},X_{i,j}|X_{S_{i,j}}\right)$ using iterated stepwise algorithms. Hence, QPC proves to be a valuable tool for variable screening within the high-dimensional quantile regression framework.
	
	Thirdly, in addition to its theoretical advantages, the QPC screening approach also offers computational benefits. When compared to commonly used penalization methods, QPC screening demonstrates superior efficiency and lower computational cost within the realm of quantile regression. We will further illustrate this computational advantage in Section 4.3.

	Shifting from the population level to sample data, we define the sample quantile partial correlation (sample QPC) denoted as $%
	\widehat{qpcor}_{\tau }$:
	\begin{equation*}
		\widehat{qpcor}_{\tau}\left(Y_{i+1},X_{i,j}|X_{i,S_{j}}\right) =\frac{\frac{1}{n}%
			\sum_{i=1}^{n}\left( \psi _{\tau }\left( Y_{i+1}-X_{i,S_{j}}^{^{T }}\hat{%
				\alpha}_{j}\right) \left( X_{i,j}-X_{i,S_{j}}^{^{T }}\hat{\theta}_{j}\right)
			\right) }{\sqrt{\tau \left( 1-\tau \right) \hat{\sigma}_{ij}^{2}}}
	\end{equation*}%
	where $\hat{\alpha}_{j}=\arg \min \frac{1}{n}\sum_{i=1}^{n}\left( \rho
	_{\tau }\left( Y_{i+1}-X_{i,S_{j}}^{^{T }}\alpha _{j}\right) \right) $, $%
	\hat{\theta}_{j}=\arg \min \frac{1}{n}\sum_{i=1}^{n}\left(
	X_{i,j}-X_{i,S_{j}}^{^{T }}\theta _{j}\right) ^{2}$, and $\hat{\sigma}%
	_{ij}^{2}=\frac{1}{n}\sum_{i=1}^{n}\left( X_{i,j}-X_{i,S_{j}}^{^{T}}\hat{%
		\theta}_{j}\right) ^{2}$. $\hat{\alpha}_{j}$, $\hat{\theta}_{j}$, and $\hat{\sigma}_{ij}^{2}$ are sample estimates of ${\alpha}^{0}_{j}$, ${\theta}^{0}_{j}$, and ${\sigma}_{ij}^{2}$.
	
	We introduce two screening algorithms for variable selection using sample data and sample QPC. These algorithms, named quantile partial correlation screening (QPCS) and quantile partial correlation forward selection (QPCFR), correspond to algorithms 1 and 3 in \citet{ma2017variable}. Both QPCS and QPCFR aim to maximize the sample QPC in each selection step. The key difference is that QPCS enlarges the conditioning set to account for potential confounding effects from other highly correlated variables. It mitigates the influence of variables in this confounding set, ensuring a more focused variable selection process. Specifically, for a given variable $X_j$, we define a confounding set: 
	\begin{align*}
		S_{j}^{\nu}\left(m_{j}\right)= & \{k\neq j\:and\:k\in S_{j}:\\
		& \left|\rho_{jk}\right|\:is\:among\:the\:first\:m_{j}\:largest\:correlation\:of\:all\:possible\:correlations\}
	\end{align*}

	Now we start to demonstrate both algorithms below. 
	
	\textbf{QPCS Algorithm:}
	
	\begin{enumerate}
		\item Initialize the active set of variables $S_1=\emptyset$.
		\item For $d = 1,\dots,d^*$, where $d^*$ is a prespecified number:
		\begin{enumerate}
			\item Given $S_{d}$, we set $\bar{S}_d = S_{d} \cup S_j^{\nu}\left(m_{j}\right)$. We find a covariate index 
			$j^{\ast }=\arg \max_{j\notin \bar{S}_{d}}\left\vert \widehat{qpcor}_{\tau}\left(Y_{i+1},X_{i,j}|X_{i,\bar{S}_d}\right) \right\vert $.
			\item Update $S_{d+1}=S_{d}\cup \left\{ j^{\ast }\right\} $
		\end{enumerate}
		\item For $d = d^*+1, \dots, D_{max}$, where $D_{max}$ is a prespecified number:
		\begin{enumerate}
			\item We set $\bar{S}_d = S_{d^*} \cup S_j^{\nu}$. We find a covariate index 
			$j^{\ast }=\arg \max_{j\notin \bar{S}_{d}}\left\vert \widehat{qpcor}_{\tau}\left(Y_{i+1},X_{i,j}|X_{i,\bar{S}_d}\right) \right\vert $.
			\item Update $S_{d+1}=S_{d}\cup \left\{ j^{\ast }\right\} $
		\end{enumerate}
		\item Given a specific choice of $D$, we obtain $\hat{\beta}_{S_{D}}=\arg \min_{\beta _{S_{D}}}%
		\frac{1}{n}\sum_{i=1}^{n}\rho _{\tau }\left( Y_{i+1}-X_{i,S_{D}}^{^{T
		}}\beta _{S_{D}}\right) $.
	\end{enumerate}
	
	\textbf{QPCFR Algorithm:}
	\begin{enumerate}
		\item Initialize the active set of variables $S_1=\emptyset$.
		
		\item For $d=1,\dots,D_{max}$:
		
		\begin{enumerate}
			\item Find a covariate index  $j^{\ast }=\arg \max_{j\notin S_{d}}\left\vert \widehat{qpcor}_{\tau}\left(Y_{i+1},X_{i,j}|X_{i,S_{d}}\right) \right\vert $
			
			\item Update $S_{d+1}=S_{d}\cup \left\{ j^{\ast }\right\} $
		\end{enumerate}
		
		\item Given a specific choice of $D$, we obtain $\hat{\beta}_{S_{D}}=\arg \min_{\beta _{S_{D}}}%
		\frac{1}{n}\sum_{i=1}^{n}\rho _{\tau }\left( Y_{i+1}-X_{i,S_{D}}^{^{T
		}}\beta _{S_{D}}\right) $.
	\end{enumerate}
	
	In the QPCS algorithm, the choice of $d^{*}$ follows \citet{ma2017variable}. We set $d^* = C_1 {\left(\frac{n}{\log{n}}\right)}^{\frac{1}{2}}$ and let $C_1 = 1$. For $m_j=|S_j^{\nu}\left(m_j\right)|$, we set $m_j \leq C_2 {\left(\frac{n}{\log{n}}\right)}^{\frac{1}{2}}$ for some constant $C_2=1$. For $D_{max}$ in both algorithms, we set $D_{max}=\left\lfloor \frac{n}{\log{n}} \right\rfloor$ following established variable screening literature. 
	
	In the final stage of both algorithms, the selection of the optimal value for $D$ becomes pivotal for obtaining the ultimate quantile regression estimates. We determine our choice of $D$ that  minimizes extended bayesian information criterion (EBIC). This strategy aligns with the two-step procedure in the variable screening literature. Initially, we employ a forward selection-type algorithm to screen variables, executing this process for a maximum of $D_{max}$ steps. Subsequently, we refine our selection set during the second step. An alternative approach at this stage involves utilizing penalization methods. However, it's worth noting that penalization methods tend to underperform when compared to the approach based on information criteria, as suggested by \citet{ma2017variable}. Consequently, we adhere to the information criteria-based approach in the final step of both algorithms to determine the optimal value for $D$.
		\\
	
	\section{Theory of QPC}
	
	In this section, we delve into two essential theoretical properties of QPC. These properties include the uniform convergence of sample QPC to the population QPC and the variable screening
	property. We follow the proof procedure outlined in \citet{ma2017variable}. Before
	extending these findings to the stationary $\beta $-mixing framework, we first give our refined version of theorem 1 in 
	\citet{ma2017variable} and restate the original one in appendix.
	
	\subsection{Theoretical properties of QPC under the i.i.d data}
	
	The uniform convergence of $\widehat{qpcor}%
	_{\tau }\left\{ Y_{i+1},X_{i,j}|X_{i,S_{j}}\right\} $ to the population QPC
	under the i.i.d framework is as follows.

	\begin{thm*}
		(Refined version of Theorem 1 in \citet{ma2017variable}) We suppose the data $\{Y_{i+1},X_i\}$ is independent and identically distributed. Under assumptions \ref{assu: C1} and \ref{assu: C2} in appendix (Assumptions (C1) and (C2) in \citet{ma2017variable}), for some universal constant $C$, $0<\kappa <\frac{1}{%
			4}$, and $r_{n}=Cn^{\omega }$ for some $0\leq \omega <2\kappa  $, we have 
		\begin{align*}
			& P\left( \sup_{1\leq j\leq p_{n}}\left\vert \widehat{qpcor}_{\tau }\left(
			Y_{i+1},X_{i,j}|X_{i,S_{j}}\right) -qpcor_{\tau }\left(
			Y_{i+1},X_{i,j}|X_{i,S_{j}}\right) \right\vert \geq Cr_{n}^{\frac{1}{2}%
			}n^{-\kappa }\right) \\
			\leq & p_{n}\left( {C}r_{n}e^{-\frac{{C}n^{1-2\kappa}}{r_{n}}%
			}+{C}e^{-{C} n^{1-4\kappa}}\right)
		\end{align*}
	\end{thm*}

	\begin{rem*}
		The theorem establishes the convergence of sample QPC to its population counterpart. We identify minor errors within the proof provided in \citet{ma2017variable}. Further details regarding these identified issues are outlined in appendix. The identified minor mistakes have implications primarily on the convergence rates presented in various theorems within their paper. For example, the second term in the right-hand side of the inequality is $e^{-Cn^{1-4\kappa}}$ instead of $r_n^2 e^{-Cn r_n^{-2}}$ in the original paper. Moreover, the range of $\kappa$ shrinks from $\left(0,\frac{1}{2}\right)$ to $\left(0,\frac{1}{4}\right)$ in order to maintain the convergence result. We address these issues later and incorporate it into our Theorem \ref{thm:1} as a special case in the following subsection.
	\end{rem*}

	\subsection{Theoretical properties of QPC under $\beta$-mixing processes}
	
	In this subsection, we first present conditions to demonstrate the uniform convergence of $\widehat{qpcor}_{\tau }\left\{
	Y_{i+1},X_{i,j}|X_{i,S_{j}}\right\} $ toward the population QPC concerning stationary $\beta$-mixing sequences. Subsequently, we present key theorems and propositions.
	
	\begin{assumption}
		\label{assu:beta-mix}The dependent sample $\left\{Y_{i+1}, X_{i,j}\right\} _{i=1}^{n}$ is
		a stationary $\beta$-mixing sequence with mixing coefficients $%
		\beta\left(n\right)\leq e^{-Bn}$ for some positive constant $B$ with every $%
		j=1,\dots,p$.\footnote{%
			The $\beta$-mixing condition implies $\alpha$-mixing (strong
			mixing) condition. Therefore, the theoretical properties of probabilistic
			bounds under $\alpha$-mixing condition from \citet{merlevede2009bernstein} hold in our proof. Moreover, the $\beta$-mixing coefficient $\beta\left(n\right)$ is defined as 
			$$\beta\left(n\right) =  \frac{1}{2} \sup \left\{
			\sum_{i=1}^{I} \sum_{j=1}^{J} \left|P\left(A_i \cap B_j \right) - P\left(A_i\right)P\left(B_j\right)  \right| 
			\right\}$$
			with $\left\{A_i
			\right\}_{i=1}^I\;is\;any\;finite\;partition\;of\;\mathcal{A}_{-\infty}^{0}$
			and
			$\left\{B_j
			\right\}_{j=1}^J\;is\;any\;finite\;partition\;of\;\mathcal{A}_{n}^{\infty}$ , where $\mathcal{A}_i^j$ is the $\sigma$-field generated by $X_i$,\dots,$X_j$ for $i<j$.
		}
	\end{assumption}
	
	\begin{assumption}
		\label{assu:bound}Suppose the data satisfy the following boundedness conditions for some positive constants $M_1,M_2,M_3$, and $M_4$, such that,  $\sup_{i,j}\left|X_{i,j}\right|\leq M_{1}$, $%
		\sup_{i,j}\left|X_{i,S_{j}}^{T}\theta_{j}^{0}\right|\leq M_{2}$, $%
		\sup_{i,j}\left|X_{i,S_{j}}^{T}\pi_{j}^{0}\right|\leq M_{3}$, and $%
		\sup_{j}\left\Vert \frac{1}{n}\sum_{i=1}^{n}X_{i,S_{j}}X_{i,j}\right\Vert
		\leq M_{4}$. for every $j=1,\dots,p$.
	\end{assumption}
	
	\begin{assumption}
		\label{assu:eigen}For every $1\leq i\leq n$ and $1\leq j\leq p$, we assume that 
		\begin{equation*}
			m\leq\lambda_{\min}\left(E\left(X_{i,S_{j}}X_{i,S_{j}}^{T}\right)\right)\leq%
			\lambda_{\max}\left(E\left(X_{i,S_{j}}X_{i,S_{j}}^{T}\right)\right)\leq M
		\end{equation*}
		for some universal positive constants $m$ and $M$.
	\end{assumption}
	
	\begin{assumption}
		\label{assu:density} For every $1\leq i\leq n$, we assume that the conditional density of $Y_{i+1}$
		on ${X_i}$, $f_{Y_{i+1}|X_i}\left(y\right)$, is 1-Lipschiz. It is bounded
		by a finite constant from above and has a lower bound from $0$. Additionally, the distribution $F_{Y_{i+1}|X_i}\left(y\right)$ is absolutely continuous in its support.
	\end{assumption}
	
	\begin{assumption}
		\label{assu:rn}There exists a sequence $r_{n}=\max_{1\leq j\leq
			p}\left|S_{j}\right|=Cn^{\gamma}$ for some $0\leq\gamma<2\kappa$ and $%
		0<\kappa<\frac{1}{4}$, serving as a control for the maximum cardinality of $S_j$.
	\end{assumption}
	
	\begin{assumption}
		\label{assu:identify} We assume $\min_{j\in
			M_{*}}\left|qpcor_{\tau}\left(Y_{i+1},X_{i,j}|X_{i,S_{j}}\right)\right|\geq
		C_{0}r_{n}^{\frac{1}{2}}n^{-\kappa}$ for some $0<\kappa<\frac{1}{4}$ and
		some universal positive constant $C_{0}$, where $M_{*}=\left\{ j:\beta_{j}^{0}\neq 0,\ 1\leq j\leq
		p\right\} $ is the set of nonzero quantile regression coefficients.
	\end{assumption}
	
	\begin{rem}
		Assumption \ref{assu:beta-mix} introduces stationary $\beta$-mixing condition to the data. Notably, this property remains preserved under various transformations of $\beta$-mixing processes, extending the applicability of our analysis. This assumption accommodates a wide range of time series models, including cases such as the $MA(\infty)$ process with exponentially decaying $MA$ coefficients or the inclusion of lagged response as explanatory variables.
		Assumption \ref{assu:bound} outlines the boundedness conditions for the predictors, resembling condition (C2) in \citet{ma2017variable}. While relaxing the boundedness condition by controlling tail behavior is theoretically possible by truncation argument, it would significantly complicate our derivations within the mixing framework. Assumption \ref{assu:eigen} places bounds on
		the maximal and minimal eigenvalues of the population Gram matrix $E X_{i,S_j}
		X_{i,S_j}^{T}$ for the stationary process $X_{i,S_j}$. Such conditions are common in the
		high-dimensional statistics or econometrics literature. Assumption \ref%
		{assu:density} ensures a regular conditional density of $Y_{i+1}$ on covariates. It prevents the situation that the density has value zero or infinity.  This type of density assumption is standard in the quantile methods. Assumption \ref{assu:rn}
		sets a rate condition for $r_n$, controlling the convergence rate of the sample QPC. This condition is readily met, as $r_n$ represents the maximum cardinality of the conditioning set. Assumption 
		\ref{assu:identify} is required for
		the variable screening consistency property. It builds a bridge between the magnitude of the population QPC and the covariate $X_j$, whose coefficient $\beta_j^{0}$ is positive. It is similar to the well-known "beta-min" condition in literature.
		The last two assumptions are adapted from assumption (C3) in
		\citet{ma2017variable}.
	\end{rem}

	Our primary goal is to screen variables by ranking their QPC values. Since we can estimate the
	sample QPC, $\widehat{qpcor_{\tau}}\left(Y_{i+1},X_{i,j}|X_{i,S_{j}}\right)$, it is crucial to establish the
	uniform convergence of $\widehat{qpcor_{\tau}}%
	\left(Y_{i+1},X_{i,j}|X_{i,S_{j}}\right)$ to the population counterpart, ${%
		qpcor_{\tau}}\left(Y_{i+1},X_{i,j}|X_{i,S_{j}}\right)$. To demonstrate this uniform convergence, we first prove the convergences results of various components of sample QPC, including $\hat{\theta}_j$, $\hat{\pi}_j$, $\frac{1}{n}\sum_{i=1}^{n}\psi_{\tau}\left(Y_{i+1}-X_{i,S_{j}}^{T}%
	\hat{\pi}\right)\left(X_{i,j}-X_{i,S_{j}}^{T}\hat{\theta}_{j}\right)$, and $%
	\hat{\sigma}_{j}^{2}$. We organize these convergence proofs  in the appendix from the lemma \ref{lem:2} to lemma \ref{lem:5}.
	
	Next, we present our theorem \ref{thm:1} below.
	
	\begin{thm}
		\label{thm:1}Under the assumption \ref{assu:beta-mix}, \ref{assu:bound}, \ref{assu:eigen},
		and \ref{assu:density}, for some positive constants $C^{*}_{1}$ and $\tilde{C}$s and a sufficiently large $n$, we have for $0<\kappa<\frac{1%
		}{4}$ and $r_{n}=\max_{1\leq j\leq p}\left|S_{j}\right|=Cn^{\gamma}$ with $%
		0\leq\gamma<2\kappa$, 
		\begin{align*}
			& P\left(\sup_{1\leq j\leq p_{n}}\left|\widehat{qpcor_{\tau}}%
			\left(Y_{i+1},X_{i,j}|X_{i,S_{j}}\right)-qpcor_{\tau}%
			\left(Y_{i+1},X_{i,j}|X_{i,S_{j}}\right)\right|\geq C^{*}_{1} r_{n}^{\frac{1}{%
					2}}n^{-\kappa}\right) \\
			\leq & p_{n} A_n 
		\end{align*}
		where $A_n = \tilde{C}_{1}r_{n}e^{-\frac{\tilde{C}_{2}n^{1-2\kappa}}{r_{n}}%
		}+\tilde{C}_{3}e^{-\tilde{C}_{4} n^{1-4\kappa}}+\tilde{C}_{5} n^{\frac{1}{2}%
		}e^{-Bn^{\frac{1}{2}}} \rightarrow 0$ as $n\rightarrow\infty$.
	\end{thm}
	
	The main procedure to establish this uniform convergence bound relies on the blocking technique, as introduced in \citet{yu1994rates} and \citet{arcones1994central}. We also make use of various concentration inequalities, with references in the proofs of lemmas. As the probabilistic bound shown above, the convergence of sample QPC to its population counterpart exhibits an exponential convergence rate. This rate holds as long as $log\left( p_n \right) = o\left( \tilde{C}_{2}\frac{n^{1-2\kappa}}{r_{n}} + \tilde{C}_{4}   n^{1-4\kappa} + Bn^{\frac{1}{2}}\right)$, which indicates the high-dimensional regime of the covariates. 
	
	It's worth noting that this uniform convergence bound, applicable under stationary $\beta$-mixing conditions, exhibits greater generality compared to the bound in the i.i.d. case. It encompasses additional terms and a possible slower convergence rate, contingent upon the interplay of variables such as $n$, $r_n$, the value of $\kappa$, and the mixing coefficient $B$. Among the three terms in $A_n$, namely $ \tilde{C}_{1}r_{n}e^{-\frac{\tilde{C}_{2}n^{1-2\kappa}}{r_{n}}}$,  $\tilde{C}_{3}e^{-\tilde{C}_{4} n^{1-4\kappa}}$, and $\tilde{C}_{5} n^{\frac{1}{2}}e^{-Bn^{\frac{1}{2}}}$, the first two also appear in the refined convergence bound for the i.i.d. case. The third term is novel, arising from our $\beta$-mixing framework. It hinges on the temporal dependence level, specifically, the parameter $B$ in the mixing coefficient. The rate of decay of this term is contingent upon the value of $B$; the larger $B$ is, the faster the third term converges to zero.  In the extreme case where $B=\infty$, third term becomes negligible and this bound reverts to the i.i.d. case.  Furthermore, this third term's presence suggests that we are not restricted to a fixed value of $B$. Instead, we can allow $B=B_n$ depending on $n$ as long as the mixing coefficient $\beta\left(n\right)=e^{-B_n n}$ follows a  geometrical decay. For instance, $B_n$ could be at least $O\left(n^{-\frac{1}{2}}\right)$ to ensure convergence. In such cases, selecting appropriate values for $r_{n}$, $B_n$, and $\kappa$ becomes more intricate since the dominant term changes as $n$ varies.  In general, if the logarithm of the number of covariates, $\log p_{n}$, exhibits the same order as above, the bound shrinks to $0$ for sufficiently large $n$.
	
	Building upon our theorem \ref{thm:1}, we can establish the variable screening consistency property, a concept widely used for variable screening methods. This property asserts that our procedure will eventually include the set of important (nonzero) coefficients.  
	To illustrate this, we define the selection set $\hat{M}_%
	{v_{n}}$ as follows:
	\begin{equation*}
		\hat{M}_{v_{n}}=\left\{ j:\widehat{qpcor}_{\tau}\left(
		Y_{i+1},X_{i,j}|X_{i,S_{j}}\right) \geq v_{n}\ for\ 1\leq j\leq p\right\}
	\end{equation*}
	where $v_{n}$ is a threshold converging to $0$. We denote the original set of covariates with nonzero coefficients as: \begin{equation*}
		M_{*}=\left\{ j:\beta_{j\tau}^{0} \neq 0\ for\ 1\leq j\leq p\right\} \end{equation*} The variable screening consistency property is established in theorem \ref{thm:2}.
	
	\begin{thm}
		\label{thm:2} Under the assumption \ref{assu:identify} and conditions of theorem %
		\ref{thm:1}, for the $\kappa$ defined above, if $%
		v_{n}=C_{1}^{*}r_{n}^{\frac{1}{2}}n^{-\kappa}$, we have 
		\begin{equation*}
			P \left(M_{*} \subset \hat{M}_{v_{n}}\right) \geq 1-s_{n}A_{n}
		\end{equation*}
		where $s_{n}=\left|M_{*}\right|$ is the number of nonzero coefficients and $A_n \rightarrow 0$ is defined in theorem \ref{thm:1}.
	\end{thm}

	Theorem 2 demonstrates that the variable screening property of QPC remains valid even when dealing with stationary $\beta$-mixing data. As the sample size approaches infinity, the probability of the significant variables being included in the selection set converges to $1$. This convergence rate inherits from theorem \ref{thm:1} and depends on our mixing parameter $B$. With this property, researchers could focus on the set $\hat{M}_{v_{n}}$ instead of all candidate predictors.
	
	Beyond the variable screening property, another desirable objective is the variable selection consistency, which implies two sets, $M_*$ and $\hat{M}_{v_{n}}$ are equal in probability. To achieve this result, we need to impose the following assumption.
	
	\begin{assumption} 
		\label{assu:sel-cons} $\min_{j\notin
			M_{*}}\left|qpcor_{\tau}\left(Y_{i+1},X_{i,j}|X_{i,S_{j}}\right)\right|\leq
		C_{1}^{*}r_{n}^{\frac{1}{2}}n^{-\kappa}$ for 
		any constant $C_{1}^{*}$, where $M_{*}=\left\{ j:\beta_{j}^{0} \neq 0,\ 1\leq j\leq
		p\right\} $.
		
	\end{assumption}
	The above assumption imposes the magnitude restrictions on the QPC of irrelevant predictors. Under this assumption, we can separate the true set of predictors from the irrelevant predictors with probability approaching to 1. Therefore, the variable selection consistency follows.
	
	\begin{proposition}
		\label{prop:1}  Under the condition of theorem 2 and assumption \ref{assu:sel-cons}, we have, as $n\rightarrow\infty$,
		\begin{equation*}
			P\left(M_{*}=\hat{M}_{v_{n}}\right)\rightarrow 1
		\end{equation*}

	\end{proposition}

	As suggested by the literature, assumption \ref{assu:sel-cons} is rather stringent as it places strict limits on the magnitude of irrelevant coefficients, which may not always be realistic. Instead, we can adopt a milder assumption below aimed at controlling the cardinality of $\hat{M}_{v_{n}}$, $\left| \hat{M}_{v_{n}} \right|$. This approach, which limits the size of our selection set, allows the variable screening method to effectively reduce the dimensionality of the data.
	
	\begin{assumption}\label{assu:qpc_control}
		For some $\zeta>0$, we assume that $\sum_{j=1}^{p} \left|qpcor_{\tau}\left(
		Y_{i+1},X_{i,j} | X_{i,S_j}
		\right)  \}\right| = O\left(n^{\zeta}\right)$.
	\end{assumption}
	This milder assumption imposes an order condition on the summation of population QPC. With our identifiability assumption \ref{assu:identify}, we can state the following proposition.
	\begin{proposition}
		\label{prop:2}  Under the condition of theorem \ref{thm:1} and assumption \ref{assu:qpc_control}, we have
		\begin{equation*}
			P\left(\left| \hat{M}_{v_{n}} \right| \leq C^{**}_1  n^{\zeta+\kappa-\frac{\gamma}{2}}\right) \geq 1 - p_{n} A_n 
		\end{equation*}
		where $0<C^{**}_1<\infty$ is a constant depending on $C^{*}_1$.
	\end{proposition}
	This proposition \ref{prop:2} provides the theoretical guarantee that the QPC screening approach can effectively reduce dimensionality even under a $\beta$-mixing framework. This implies its potential applicability in handling high-dimensional macroeconomic and financial datasets. Furthermore, the probability of convergence is influenced by the mixing coefficient $B$, as discussed earlier.
	
	Typically, variable screening is a two-step procedure that narrows down variable selection twice to obtain the best results. In our subsequent simulations and applications, we employ the EBIC to make the final variable selection in the second step. With proposition \ref{prop:2} and the stated conditions, we can establish screening consistency using the two-step EBIC approach, i.e., $P\left(M_{*} \subset \hat{M}_{EBIC}\right) \rightarrow 1$, where $\hat{M}_{EBIC}$ represents the selection set from the two-step EBIC procedure. The concept and proof of the screening consistency under EBIC closely resemble theorem 3 in \citet{ma2017variable}. Readers can refer to their Section 4.2 for further details.

	\section{Monte Carlo simulations}
	
	In this section, we conduct Monte Carlo simulations to evaluate the effectiveness of the QPC method. We examine the performance of the QPC screening method using both QPCS and QPCFR algorithms mentioned earlier. We also compare the performance with the commonly employed $l_1$-penalized quantile regression, focusing on out-of-sample prediction and the selection of significant variables.  For the QPC approaches, we adhere to the practices outlined in the screening literature, involving a two-stage selection process.
	
	The primary motivation behind employing a two-stage selection procedure lies in the fact that variable screening consistency guarantees the inclusion of the true non-zero coefficient set in the selected set with a probability approaching one. However, it does not guarantee an exact match between these two sets in finite samples, especially in the absence of our assumption \ref{assu:sel-cons}. In the two-stage selection process, we first employ QPC to screen variables and refine our selection criteria across covariates based on the screening consistency property. Subsquently, we use EBIC to finalize our choice of variables. This procedure is widely adopted in the variable screening literature when dealing with (ultra-)high-dimensional covariates, as seen in studies such as \citet{ma2017variable}, \citet{yousuf2018variable}, and others. Another  straightforward advantage of this procedure is its computational efficiency, which we will address in the subsection 4.3. 
	
	For both two-step QPC screening and $l_1$-penalization approaches, we employ the same EBIC to select the appropriate tuning parameters. In \citet{ma2017variable}, their numerical results indicate that a type of EBIC, introduced by \citet{lee2014model}, yields the best simulation performance. Consequently, we utilize this EBIC consistently throughout our simulations.
	$$EBIC:\quad \ln\left(\frac{1}{n}\sum_{t=1}^{n}\rho\left(Y_{t}-X_{t}^{T}\hat{\beta}_{\tau}\right)\right)+D \frac{\log n}{2n}\log D$$
	In the EBIC approach, the tuning parameter $D$ corresponds to the number of variables selected in the solution path. The solution path is determined by the number of steps in QPC approaches, whereas in the $l_{1}$-penalization approach, it depends on the penalty term $\lambda$. In both methods, we select the optimal $D\in\left\{ 1,\dots,D_{max}\right\} $
	by minimizing the EBIC criteria.

	In our simulation, we fix the setup as $n=200$, $p=1000$ and the number of nonzero coefficients $s=4$. We repeat this simulation procedure $50$ times. To assess the performance of various methods, we use the following criteria.
	
	\begin{enumerate}
		\item MQE:  mean quantile prediction error over 10 out-of-sample periods across simulations.
		\item Crate: correct selection times for all simulations. ($0\sim50$)
		\item TP: average number of true positive covariates across simulations. ($0\sim4$)
		\item FP: average number of false positive covariates across simulations. 
		\item $R_{i}$ for $i=1,2,3,4$: the average selection rank of covariate $X_i$. $R_{i}=\text{"NA"}$ means $X_i$ is not always selected, so the average rank is not available. For the $l_1$-QR, $R_{i}= \text{"-"} $ means not applicable since $l_1$-QR is not a stepwise selection method.
	\end{enumerate}
	The above criteria allow us to assess performance from different perspectives. "MQE" measures prediction error, "Crate" indicates how often the true set is correctly specified (related to selection consistency), "TP" is the most important one for variable screening consistency, and "FP" evaluates the control of the selection set's cardinality.  The $R_i$ values represent the selection rank, indicating how early important variables are selected. Ideally, we prefer ``TP'' to be $4$, which confirms the variable screening consistency, and a smaller "FP" to control the selection of irrelevant covariates. For the simulation design, we generate the linear models considered by \citet{ma2017variable} and \citet{yousuf2018variable} in the subsequent subsections.

	\subsection{DGP modified from \citet{ma2017variable}}
	
	Consider the following DGP for every time period $t$
	\begin{align*}
		y_{t} & =\beta x_{1t}+\beta x_{2t}+\beta x_{3t}-3\sqrt{\rho}\beta x_{4t}+ \sum_{l=5}^{p} 0 x_{lt} + \varepsilon_{t}
	\end{align*}
	and the predictors $X_{t}$ follow the stationary VAR(1) process
	\[
	X_{t}=\Phi X_{t-1}+\eta_{t}
	\]
	where $\Phi = diag\{\phi, \dots, \phi \}$. The error term $\varepsilon_{t}\sim z_{t}-\sigma\Psi^{-1}\left(\tau\right)$,
	and $z_{t}\sim N\left(0,\sigma^{2}\right)$ where $\Psi^{-1}\left(\cdot\right)$
	is the inverse of the standard normal cumulative density function with
	$\sigma=1$. For the innovation $\eta_{t}$ in the VAR(1) process, we define $\eta_{t}\sim N\left(0,\Sigma_{\eta}\right)$ where
	$\Sigma_{\eta}=\left\{ \sigma_{\eta_{ij}}\right\} $ is a $p\times p$
	covariance matrix with the following properties: $\sigma_{\eta_{ii}}=1$
	, $\sigma_{\eta_{ij}}=\rho\left(1-\phi^{2}\right)$ for all $i\neq j$
	and $i,j\neq4$, and $\sigma_{\eta_{4j}}=\sigma_{\eta_{i4}}=\sqrt{\rho}\left(1-\phi^{2}\right)$
	in order to rule out the marginal correlation between $x_{4t}$ and
	$y_{t}$ in population. For the coefficients, we assign $\beta=2.5\left(1+\left|\tau-0.5\right|\right)$
	as \citet{ma2017variable}. 
	
	Under this setup, we report the results for various $\tau=0.2,0.5,0.8$ to accommodate different quantile levels.Additionally, we have explored different choices for $\phi$ with $\phi=0.2, 0.5, 0.8$, which represent different levels of data persistence. Furthermore, we have considered two different values for $\rho$, specifically $\rho=0.5$ and $\rho=0.95$, to introduce moderate and high correlation among the covariates.  These variations in parameters allow us to comprehensively evaluate the performance of the QPC screening method.

	In tables 1-3, we present a performance comparison of the QPCS, QPCFR, and $l_{1}$-QR procedures under the condition $\rho=0.5$. When the time series dependence $\phi$ is low, both QPCS and QPCFR methods include all $4$ important variables in the screening step. This confirms the variable screening consistency property. Additionally, we observe that the screening approaches outperform the $l_{1}$-QR method.  Specifically, under the cases of $\phi=0.2,0.5$, the numbers of false positive covariates is very low (less than 2) across all simulations, and the exact selection rate exceeds 50\% ($\geq25$) for QPCS, especially at $\tau=0.5$. In contrast, the $l_{1}$ penalization method tends to select more irrelevant predictors. However, when the dependence level is high ($\phi=0.8$), the performance of QPCS deteriorates rapidly, while QPCFR continues to achieve promising results. This discrepancy with the behavior observed in i.i.d. data, where \citet{ma2017variable} suggested that QPCS outperforms QPCFR, is noteworthy. Moreover, when examining the average selection rank for the first four covariates, our findings reveal that both QPCS and QPCFR consistently select the significant covariates at the early stage for the cases with $\phi=0.2$ and $\phi=0.5$. In scenarios with $\phi=0.8$, although QPCS cannot consistently include the correct covariates, resulting in a "NA" average selection rank in the table, QPCFR still reports a low average selection rank. Regarding our benchmark $l_{1}$-QR method, it is inappropriate to assess the selection rank for this penalization approach, denoted as "-" in the tables. Overall, two screening methods outperform $l_{1}$-QR in terms of out-of-sample prediction under $\phi=0.2$ and $\phi=0.5$. In summary, our simulations provide strong evidence that the two-step QPC with EBIC is a valuable tool for variable screening/selection in high-dimensional time-series data.

	\begin{table}[H]
			\small
		\caption{Performance comparison of two QPC screening procedures and $l_{1}$-QR
			method, \emph{$\rho=0.5$}, $\phi=0.2$ \label{tab:comp_rho0.5phi0.2}}
	
		\begin{centering}
			\begin{tabular}{cccccccccc}
				\hline 
				$\tau$ & Method & MQE & Crate & TP & FP & $R_{1}$ & $R_{2}$ & $R_{3}$ & $R_{4}$\tabularnewline
				\hline 
				& QPCS & 0.3073  & 31 & 4 & 0.6 & 2.02 & 1.96 & 2.02 & 4\tabularnewline
				0.2 & QPCFR & 0.3053  & 4 & 4 & 2.26 & 3.28 & 3 & 2.88 & 3.4\tabularnewline
				& $l_{1}$-QR & 1.0396  & 0 & 2.76 & 8.78 & - & - & - & -\tabularnewline
				\hline 
				& QPCS & 0.4258  & 27 & 4 & 0.64 & 2.14 & 1.86 & 2 & 4\tabularnewline
				0.5 & QPCFR & 0.4284  & 11 & 4 & 1.24 & 3.18 & 2.2 & 2.96 & 2.58\tabularnewline
				& $l_{1}$-QR & 1.2485  & 0 & 2.18 & 2.82 & - & - & - & -\tabularnewline
				\hline 
				& QPCS & 0.3032  & 19 & 4 & 0.92 & 2.06 & 1.94 & 2.02 & 3.98\tabularnewline
				0.8 & QPCFR & 0.3072  & 4 & 4 & 2.06 & 3.22 & 2.62 & 2.98 & 3.08\tabularnewline
				& $l_{1}$-QR & 1.1061  & 0 & 2.84 & 9.18 & - & - & - & -\tabularnewline
				\hline 
			\end{tabular}
			\par\end{centering}
			\footnotesize
		\begin{note*}
			This table summarizes the results generated by DGP in section 4.1 across 50 simulations.
			The simulation setup is as follows: $n=200$, $p=1000$, and $\phi=0.2$. We evaluate
			the performance of QPCS, QPCFR, and $l_{1}$-QR across different quantile
			levels $\tau\in\left\{ 0.2,0.5,0.8\right\} $.
		\end{note*}
	\end{table}
	
	\begin{table}[H]
		\caption{Performance comparison of two QPC screening procedures and $l_{1}$-QR
			method, \emph{$\rho=0.5$,} $\phi=0.5$ \label{tab:comp_rho0.5phi0.5}}
		\small
		\begin{centering}
			\begin{tabular}{cccccccccc}
				\hline 
				$\tau$ & Method & MQE & Crate & TP & FP & $R_{1}$ & $R_{2}$ & $R_{3}$ & $R_{4}$\tabularnewline
				\hline 
				& QPCS & 0.3073  & 27 & 4 & 0.74 & 2.78 & 2.82 & 2.72 & 2.24\tabularnewline
				0.2 & QPCFR & 0.3181  & 8 & 4 & 1.84 & 2.8 & 2.78 & 3.04 & 1.92\tabularnewline
				& $l_{1}$-QR & 0.4487  & 4 & 4 & 3.26 & - & - & - & -\tabularnewline
				\hline 
				& QPCS & 0.4228  & 32 & 4 & 0.5 & 2.6 & 2.62 & 2.64 & 2.14\tabularnewline
				0.5 & QPCFR & 0.4272  & 17 & 4 & 0.96 & 2.88 & 2.56 & 2.78 & 1.78\tabularnewline
				& $l_{1}$-QR & 0.5892  & 3 & 4 & 2.74 & - & - & - & -\tabularnewline
				\hline 
				& QPCS & 0.3005  & 30 & 4 & 0.52 & 2.74 & 2.7 & 2.66 & 2.06\tabularnewline
				0.8 & QPCFR & 0.3000  & 11 & 4 & 1.46 & 3.02 & 2.8 & 3.06 & 1.9\tabularnewline
				& $l_{1}$-QR & 0.4823  & 4 & 4 & 3.74 & - & - & - & -\tabularnewline
				\hline 
			\end{tabular}
			\par\end{centering}
			\footnotesize
		\begin{note*}
			This table summarizes the results generated by DGP in section 4.1 across 50 simulations.
				The simulation setup is as follows:
				$n=200$, $p=1000$, $\rho=0.5$, and $\phi=0.5$.
				We evaluate the performance of QPCS, QPCFR, and $l_{1}$-QR across
				different quantile levels $\tau\in\left\{ 0.2,0.5,0.8\right\} $.
		\end{note*}
	\end{table}
	
	\begin{table}[H]
		\caption{Performance comparison of two QPC screening procedures and $l_{1}$-QR
			method, \emph{$\rho=0.5$,} $\phi=0.8$ \label{tab:comp_rho0.5phi0.8}}
		\small
		\begin{centering}
			\begin{tabular}{cccccccccc}
				\hline 
				$\tau$ & Method & MQE & Crate & TP & FP & $R_{1}$ & $R_{2}$ & $R_{3}$ & $R_{4}$\tabularnewline
				\hline 
				& QPCS & 1.4182  & 5 & 3.14 & 6.34 & NA & NA & NA & NA\tabularnewline
				0.2 & QPCFR & 0.2933  & 1 & 4 & 2.44 & 4.52 & 4.44 & 4.7 & 2.52\tabularnewline
				& $l_{1}$-QR & 0.8475  & 0 & 4 & 5.52 & - & - & - & -\tabularnewline
				\hline 
				& QPCS & 0.5337  & 25 & 3.88 & 1.02 & 4.3 & NA & NA & 1.18\tabularnewline
				0.5 & QPCFR & 0.4352  & 25 & 4 & 0.72 & 3.06 & 2.96 & 3.14 & 1.24\tabularnewline
				& $l_{1}$-QR & 0.7976  & 0 & 4 & 5.52 & - & - & - & -\tabularnewline
				\hline 
				& QPCS & 2.1786  & 2 & 3.1 & 5.1 & NA & NA & NA & NA\tabularnewline
				0.8 & QPCFR & 0.3083  & 6 & 4 & 1.98 & 4.18 & 3.78 & 4.44 & 2.22\tabularnewline
				& $l_{1}$-QR & 0.7535  & 0 & 4 & 6.06 & - & - & - & -\tabularnewline
				\hline 
			\end{tabular}
			\par\end{centering}
			\footnotesize
		\begin{note*}
			This table summarizes the results generated by DGP in section 4.1 across 50 simulations. The simulation setup is as follows: $n=200$, $p=1000$, $\rho=0.5$, and $\phi=0.8$.
				We evaluate the performance of QPCS, QPCFR, and $l_{1}$-QR across
				different quantile levels $\tau\in\left\{ 0.2,0.5,0.8\right\} $.
		\end{note*}
	\end{table}
	
	In tables 4-6, we present the results for simulations with $\rho=0.95$. The performance metrics are similar to those observed in the cases with $\rho=0.5$. This consistency in performance across different correlation levels highlights the robustness of QPC methods.
	
	\begin{table}[H]
		\caption{Performance comparison of two QPC screening procedures and $l_{1}$-QR
			method, \emph{$\rho=0.95$,} $\phi=0.2$ \label{tab:comp_rho0.95phi0.2}}
		\small
		\begin{centering}
			\begin{tabular}{cccccccccc}
				\hline 
				$\tau$ & Method & MQE & Crate & TP & FP & $R_{1}$ & $R_{2}$ & $R_{3}$ & $R_{4}$\tabularnewline
				\hline 
				& QPCS & 0.3047  & 26 & 4 & 0.78 & 2.8 & 3.1 & 2.66 & 1.44\tabularnewline
				0.2 & QPCFR & 0.3107  & 11 & 4 & 1.44 & 2.88 & 2.7 & 2.9 & 1.6\tabularnewline
				& $l_{1}$-QR & 0.3943  & 7 & 4 & 3.18 & - & - & - & -\tabularnewline
				\hline 
				& QPCS & 0.4256  & 32 & 4 & 0.46 & 2.92 & 2.88 & 3.04 & 1.16\tabularnewline
				0.5 & QPCFR & 0.4364  & 15 & 4 & 1.04 & 2.68 & 2.56 & 3.28 & 1.56\tabularnewline
				& $l_{1}$-QR & 0.5298  & 5 & 4 & 2.7 & - & - & - & -\tabularnewline
				\hline 
				& QPCS & 0.3041  & 31 & 4 & 0.56 & 2.9 & 2.98 & 2.86 & 1.32\tabularnewline
				0.8 & QPCFR & 0.3066  & 11 & 4 & 1.6 & 3 & 2.72 & 2.94 & 1.56\tabularnewline
				& $l_{1}$-QR & 0.4177  & 7 & 4 & 3.3 & - & - & - & -\tabularnewline
				\hline 
			\end{tabular}
			\par\end{centering}
			\footnotesize
		\begin{note*}
			This table summarizes the results generated by DGP in section 4.1 across 50 simulations.
				The simulation setup is as follows: $n=200$, $p=1000$, $\rho=0.95$, and $\phi=0.2$.
				We evaluate the performance of QPCS, QPCFR, and $l_{1}$-QR across
				different quantile levels $\tau\in\left\{ 0.2,0.5,0.8\right\} $.
		\end{note*}
	\end{table}
	
	\begin{table}[H]
		\caption{Performance comparison of two QPC screening procedures and $l_{1}$-QR
			method, \emph{$\rho=0.95$,} $\phi=0.5$ \label{tab:comp_rho0.95phi0.5}}
		\small
		\begin{centering}
			\begin{tabular}{cccccccccc}
				\hline 
				$\tau$ & Method & MQE & Crate & TP & FP & $R_{1}$ & $R_{2}$ & $R_{3}$ & $R_{4}$\tabularnewline
				\hline 
				& QPCS & 0.3043  & 27 & 4 & 0.68 & 3.16 & 3.12 & 3.26 & 1\tabularnewline
				0.2 & QPCFR & 0.3084  & 9 & 4 & 1.74 & 3.44 & 3.44 & 3.42 & 1.36\tabularnewline
				& $l_{1}$-QR & 0.4129  & 3 & 4 & 4.82 & - & - & - & -\tabularnewline
				\hline 
				& QPCS & 0.4168  & 38 & 4 & 0.36 & 2.96 & 3.08 & 2.96 & 1\tabularnewline
				0.5 & QPCFR & 0.4236  & 15 & 4 & 1.12 & 3.22 & 2.84 & 3.02 & 1.14\tabularnewline
				& $l_{1}$-QR & 0.5460  & 2 & 4 & 4.5 & - & - & - & -\tabularnewline
				\hline 
				& QPCS & 0.3034  & 32 & 4 & 1.2 & 3.3 & 3.28 & 3.64 & 1\tabularnewline
				0.8 & QPCFR & 0.3070  & 9 & 4 & 1.58 & 3.36 & 3.1 & 3.4 & 1.2\tabularnewline
				& $l_{1}$-QR & 0.4215  & 0 & 4 & 5.3 & - & - & - & -\tabularnewline
				\hline 
			\end{tabular}
			\par\end{centering}
			\footnotesize
		\begin{note*}
			This table summarizes the results generated by DGP in section 4.1 across 50 simulations.
				The simulation setup is as follows: $n=200$, $p=1000$, $\rho=0.95$, and $\phi=0.5$.
				We evaluate the performance of QPCS, QPCFR, and $l_{1}$-QR across
				different quantile levels $\tau\in\left\{ 0.2,0.5,0.8\right\} $.
		\end{note*}
	\end{table}
	
	\begin{table}[H]
		\caption{Performance comparison of two QPC screening procedures and $l_{1}$-QR
			method, \emph{$\rho=0.95$,} $\phi=0.8$ \label{tab:comp_rho0.95phi0.8}}
		\small
		\begin{centering}
			\begin{tabular}{cccccccccc}
				\hline 
				$\tau$ & Method & MQE & Crate & TP & FP & $R_{1}$ & $R_{2}$ & $R_{3}$ & $R_{4}$\tabularnewline
				\hline 
				& QPCS & 2.0018  & 2 & 2.86 & 7.16 & NA & NA & NA & 1.84\tabularnewline
				0.2 & QPCFR & 0.3045  & 3 & 4 & 2.16 & 4.36 & 4.06 & 4.14 & 1.52\tabularnewline
				& $l_{1}$-QR & 0.9243  & 0 & 4 & 6.32 & - & - & - & -\tabularnewline
				\hline 
				& QPCS & 0.7652  & 23 & 3.64 & 2.18 & NA & NA & NA & 1\tabularnewline
				0.5 & QPCFR & 0.4304  & 26 & 4 & 0.64 & 3.22 & 3.04 & 3.26 & 1.06\tabularnewline
				& $l_{1}$-QR & 0.8984  & 0 & 3.96 & 5.98 & - & - & - & -\tabularnewline
				\hline 
				& QPCS & 2.1604  & 3 & 3 & 6.5 & NA & NA & NA & 1.42\tabularnewline
				0.8 & QPCFR & 0.3099  & 6 & 4 & 1.92 & 4.22 & 4.04 & 4.14 & 1.46\tabularnewline
				& $l_{1}$-QR & 0.8010  & 0 & 3.98 & 6.76 & - & - & - & -\tabularnewline
				\hline 
			\end{tabular}
			\par\end{centering}
			\footnotesize
		\begin{note*}
			This table summarizes the results generated by DGP in section 4.1 across 50 simulations.
				The simulation setup is as follows: $n=200$, $p=1000$, $\rho=0.95$, and $\phi=0.8$.
				We evaluate the performance of QPCS, QPCFR, and $l_{1}$-QR across
				different quantile levels $\tau\in\left\{ 0.2,0.5,0.8\right\} $.
		\end{note*}
	\end{table}

	\subsection{DGP modified from \citet{yousuf2018variable}}
	
	We consider the following DGP for every time period $t$
	\begin{align*}
		y_{t} & =\sum_{i=1}^{p}\beta_{i}x_{it}+\varepsilon_{t}
	\end{align*}
	and the predictors $X_{t}=\left(x_{1t},\dots,x_{pt}\right)^{T}$ follow
	the stationary VAR(1) processes
	
	\[
	X_{t}=\Phi X_{t-1}+\eta_{t}
	\]
	where $\Phi=diag \{\phi,\dots,\phi  \}$. We model $\varepsilon_{t}$ as $\varepsilon_{t}\sim z_{t}-\sigma\Psi^{-1}\left(\tau\right)$,
	and $z_{t}\sim N\left(0,\sigma^{2}\right)$ where $\Psi^{-1}\left(\cdot\right)$
	is the inverse of standard normal cumulative density function with
	$\sigma=1$. The innovation $\eta_{t}$ in the VAR(1) process of $X_{t}$ is drawn from $\eta_{t}\sim N\left(0,\Sigma_{\eta}\right)$ where
	$\Sigma_{\eta}=\left\{ \sigma_{\eta_{ij}}\right\} =\left\{ \rho^{\left|i-j\right|}\right\} $
	is a $p\times p$ covariance matrix of $\eta_{t}$. The coefficients $\beta_{i}$ are set as $\beta_{i}=1$ for $i=1,2,3,4$, and $\beta_{i}=0$ for $i>4$. We report the simulation results in the tables 7-9 below for different quantile levels $\tau=0.2,0.5,0.8$ and varying persistence levels represented by $\phi=0.2,0.5,0.8$. The correlation among the covariates is set at $\rho=0.5$.
	
	\begin{table}[H]
		\caption{Performance comparison of two QPC screening procedures and $l_{1}$-QR
			method, \emph{$\rho=0.5$,} $\phi=0.2$ \label{tab:comp_rho0.5phi0.2_yousuf}}
		\small
		\begin{centering}
			\begin{tabular}{cccccccccc}
				\hline 
				$\tau$ & Method & MQE & Crate & TP & FP & $R_{1}$ & $R_{2}$ & $R_{3}$ & $R_{4}$\tabularnewline
				\hline 
				& QPCS & 0.2994  & 36 & 4 & 0.58 & 2.32 & 2.62 & 2.2 & 2.86\tabularnewline
				0.2 & QPCFR & 0.3100  & 7 & 4 & 1.76 & 2.72 & 2.32 & 2.1 & 2.86\tabularnewline
				& $l_{1}$-QR & 0.3275  & 36 & 4 & 0.42 & - & - & - & -\tabularnewline
				\hline 
				& QPCS & 0.4182  & 35 & 4 & 0.38 & 2.56 & 2.22 & 1.98 & 3.24\tabularnewline
				0.5 & QPCFR & 0.4374  & 16 & 4 & 0.98 & 2.78 & 2.22 & 2.14 & 2.86\tabularnewline
				& $l_{1}$-QR & 0.4403  & 37 & 4 & 0.34 & - & - & - & -\tabularnewline
				\hline 
				& QPCS & 0.2960  & 34 & 4 & 0.42 & 2.38 & 2.14 & 2.5 & 2.98\tabularnewline
				0.8 & QPCFR & 0.3032  & 8 & 4 & 1.8 & 2.76 & 2.22 & 2.32 & 2.7\tabularnewline
				& $l_{1}$-QR & 0.3241  & 30 & 4 & 0.56 & - & - & - & -\tabularnewline
				\hline 
			\end{tabular}
			\par\end{centering}
			\footnotesize
		\begin{note*}
			This table summarizes the results generated by DGP in section 4.2 across 50 simulations.
				The simulation setup is as follows: $n=200$, $p=1000$, $\rho=0.5$, and $\phi=0.2$.
				We evaluate the performance of QPCS, QPCFR, and $l_{1}$-QR across
				different quantile levels $\tau\in\left\{ 0.2,0.5,0.8\right\} $.
		\end{note*}
	\end{table}
	
	\begin{table}[H]
		\caption{Performance comparison of two QPC screening procedures and $l_{1}$-QR
			method, \emph{$\rho=0.5$,} $\phi=0.5$ \label{tab:comp_rho0.5phi0.5_yousuf}}
		\small
		\begin{centering}
			\begin{tabular}{cccccccccc}
				\hline 
				$\tau$ & Method & MQE & Crate & TP & FP & $R_{1}$ & $R_{2}$ & $R_{3}$ & $R_{4}$\tabularnewline
				\hline 
				& QPCS & 0.3036  & 29 & 4 & 0.64 & 2.4 & 2.62 & 2.28 & 3.22\tabularnewline
				0.2 & QPCFR & 0.3113  & 9 & 4 & 1.58 & 2.6 & 2.5 & 1.88 & 3.02\tabularnewline
				& $l_{1}$-QR & 0.3338  & 26 & 4 & 0.62 & - & - & - & -\tabularnewline
				\hline 
				& QPCS & 0.4188  & 40 & 4 & 0.2 & 2.56 & 2.38 & 2.14 & 2.92\tabularnewline
				0.5 & QPCFR & 0.4332  & 15 & 4 & 0.98 & 2.82 & 2.12 & 2.36 & 2.7\tabularnewline
				& $l_{1}$-QR & 0.4418  & 34 & 4 & 0.42 & - & - & - & -\tabularnewline
				\hline 
				& QPCS & 0.2987  & 35 & 4 & 0.5 & 2.66 & 2.36 & 2.34 & 2.94\tabularnewline
				0.8 & QPCFR & 0.3100  & 17 & 4 & 1.3 & 2.78 & 2.24 & 2.22 & 2.76\tabularnewline
				& $l_{1}$-QR & 0.3321  & 25 & 4 & 0.62 & - & - & - & -\tabularnewline
				\hline 
			\end{tabular}
			\par\end{centering}
			\footnotesize
		\begin{note*}
			This table summarizes the results generated by DGP in section 4.2 across 50 simulations.
				The simulation setup is as follows: $n=200$, $p=1000$, $\rho=0.5$, and $\phi=0.5$.
				We evaluate the performance of QPCS, QPCFR, and $l_{1}$-QR across
				different quantile levels $\tau\in\left\{ 0.2,0.5,0.8\right\} $.
		\end{note*}
	\end{table}
	
	\begin{table}[H]
		\caption{Performance comparison of two QPC screening procedures and $l_{1}$-QR
			method, \emph{$\rho=0.5$,} $\phi=0.8$ \label{tab:comp_rho0.5phi0.8_yousuf}}
		\small
		\begin{centering}
			\begin{tabular}{cccccccccc}
				\hline 
				$\tau$ & Method & MQE & Crate & TP & FP & $R_{1}$ & $R_{2}$ & $R_{3}$ & $R_{4}$\tabularnewline
				\hline 
				& QPCS & 0.3297  & 15 & 3.76 & 2.18 & NA & NA & 3.98 & NA\tabularnewline
				0.2 & QPCFR & 0.3029  & 15 & 4 & 1.28 & 2.68 & 2.4 & 2.26 & 3.06\tabularnewline
				& $l_{1}$-QR & 0.3627  & 20 & 4 & 1.02 & - & - & - & -\tabularnewline
				\hline 
				& QPCS & 0.4152  & 43 & 4 & 0.2 & 2.68 & 2.28 & 2.16 & 3.12\tabularnewline
				0.5 & QPCFR & 0.4316  & 23 & 4 & 0.62 & 2.82 & 2.18 & 2.1 & 2.9\tabularnewline
				& $l_{1}$-QR & 0.4784  & 23 & 4 & 0.74 & - & - & - & -\tabularnewline
				\hline 
				& QPCS & 0.2980  & 16 & 3.88 & 2.52 & NA & NA & 3.76 & 5.64\tabularnewline
				0.8 & QPCFR & 0.3013  & 19 & 4 & 0.9 & 2.92 & 2.22 & 2.4 & 2.74\tabularnewline
				& $l_{1}$-QR & 0.3520  & 23 & 4 & 0.86 & - & - & - & -\tabularnewline
				\hline 
			\end{tabular}
			\par\end{centering}
			\footnotesize
		\begin{note*}
			This table summarizes the results generated by DGP in section 4.2 across 50 simulations.
				The simulation setup is as follows: $n=200$, $p=1000$, $\rho=0.5$, and $\phi=0.8$.
				We evaluate the performance of QPCS, QPCFR, and $l_{1}$-QR across
				different quantile levels $\tau\in\left\{ 0.2,0.5,0.8\right\} $.
		\end{note*}
	\end{table}
	
	The simulation results indicate that both screening methods outperform the $l_1$-QR method in terms of prediction performance across all scenarios. When the time series dependency level is low, such as $\phi = 0.2$ and $\phi = 0.5$, both screening methods successfully achieve variable screening consistency, as indicated by TP=4. Additionally, the correct specification rate of QPCS is quite high and similar to that of the penalization method. For the selection rank of important variables, the average rank is around 2, suggesting that all important variables are consistently selected at the early stages of the algorithms. 
	
	However, scenarios with $\phi=0.8$, the QPCFR method consistently delivers excellent results, while the performance of the QPCS method deteriorates, as it struggles to consistently select the correct variables. These findings underscore the importance of careful algorithm selection when dealing with data that exhibits high level of time series dependency.

	\subsection{Advantage of screening methods on computational efficiency}
	
	Using variable screening methods offers a significant advantage in terms of computational efficiency when compared to penalization methods like $l_1$-penalized quantile regression. This efficiency becomes particularly evident when searching for the optimal tuning parameter.
	
	To find the optimal tuning parameter, a common approach is grid search, where an interval of $\lambda$ values is divided into numerous partitions, and the optimal $\lambda$ is determined by minimizing certain information criteria. This process often involves repeatedly solving the optimization problem, which can be especially time-consuming when applied to quantile regression models.
	
	In contrast, variable screening approaches rely on forward selection-type algorithms. Researchers progress through steps such as $1,2,3,\dots$, and the tuning parameter is the number of steps, denoted as $K$. If the model is relatively sparse, researchers typically do not need to explore an extensive range of values for $K$ in practice. Moreover, the computational time spent repeatedly solving quantile regressions is significantly less than that required for optimizing penalized objective functions.
	
	The following table provides a straightforward comparison of the computational costs between QPCS, QPCFR, and $l_1$-QR. We conducted a numerical simulation as described in Section 4.1 with $n=100$, $p=500$, and a total of 100 simulations. Furthermore, we maintained a fixed set of DGPs for each method. Our results clearly demonstrate that the running speeds of both screening approaches are more than 20 times faster than the $l_1$-penalization method.
	
	\begin{table}[H]
		\caption{Computation cost comparison across $S=100$ simulations \label{tab:computation}}
		\small
		\begin{centering}
			\begin{tabular}{cc}
				\hline 
				Method & Average running time (sec.)\tabularnewline
				\hline 
				QPCS & 16.01 \tabularnewline
				QPCFR & 15.86\tabularnewline
				$l_{1}$-QR & 344.60\tabularnewline
				\hline 
			\end{tabular}
			\par\end{centering}
			\footnotesize
		\begin{note*}
			This table presents the average computational time for 1 simulation of the QPCS, QPCFR, and $l_{1}$-QR methods. The computational time is calculated as the total running time divided by the number of simulations, which is $S=100$, and the unit is seconds.
		\end{note*}
	\end{table}
	
	\section{Macroeconomic variable screening for growth-at-risk prediction}
	
	Measuring and predicting the downside risk of economic growth has gained increasing attention.  \citet{adrian2019vulnerable} introduce the growth-at-risk (GaR) metric to evaluate this risk, estimating the conditional quantile of the real GDP growth rate at the 5\% level. Since its inception, several researchers have enhanced GaR forecasting.  Debates have arisen regarding the predictability of GaR and the economic indicators that can effectively forecast it. Initially, \citet{adrian2019vulnerable} employed the National Financial Condition Index (NFCI) as the primary predictor, asserting that a tighter financial condition increases downside risk They also elucidated the evolution of GaR. Subsequently, \citet{brownlees2021backtesting} conducted backtesting of GaR forecasting using quantile regressions and GARCH models. Their findings suggested that standard volatility models like GARCH can yield more accurate predictions. Another approach considered by \citet{plagborg2020growth} involves utilizing extracted factors to predict GaR. They argued that the forecasting capability of financial variables is limited in this context.
	
	In this section, we adopt a different perspective to address the problem. While previous literature often relies on a limited set of indexes or factors for GaR forecasting, we emphasize data-driven macroeconomic variable selection. Our central question revolves around identifying the critical macroeconomic variables that influence GaR prediction. To tackle this question, we employ our QPC technique, leveraging a substantial number of macroeconomic predictors.
	
	From a theoretical perspective, QPC exhibits variable screening property, making it a reliable tool for determining which variables should be included in the model. This property is particularly effective in the context of high-dimensional time series data, rendering it suitable for large macroeconomic datasets. Furthermore, our variable screening approach allows for the ranking of the relative importance among these variables. As previously discussed in Section 2, the QPC value for each variable is closely tied to the corresponding coefficient in multivariate quantile regression. This implies that if a variable is selected by our QPC procedure (i.e., $qpcor_{\tau}\left(Y_{i+1},X_{i,j}|X_{i,S_j}\right) \neq 0$), it signifies that this variable has a non-zero impact on GaR prediction. Moreover, since QPC quantifies partial correlation, it effectively isolates the influence originating from other predictors, providing a precise assessment of a variable's significance in GaR prediction. 
	
	From an empirical standpoint, relying solely on indexes or factors may result in information loss. In contrast, the QPC approach rigorously evaluates each variable, enabling us to extract the maximum information from the dataset.

	In our empirical analysis, we employ a truncated dataset that comprises the U.S. GDP growth rate and 243 predictors sourced from Fred-QD. This dataset has been made available by \citet{mccracken2020fred}.  It is a quarterly dataset encompassing U.S. macroeconomic time series. A distinguishing feature of this dataset is its extensive collection of indicators, albeit over a relatively short time span. Our selected time frame spans from the third quarter of 1987 (1987Q3) to the third quarter of 2022 (2022Q3), containing a total of 141 quarters.  We we focus on one-step ahead prediction utilizing quantile regression, represented as follows:
	$$Q_{\tau}\left(Y_{t+1}|X_{t}\right)=\beta_{1}X_{1t}+\cdots+\beta_{p}X_{pt}$$
	with $\tau=0.05$, $Y_{t}$ representing the U.S. GDP growth rate, and $X_{t}$ indicating predictors. 
	
	In the initial step, we adhere to the guidelines provided in \citet{mccracken2016fred} by transforming all variables into stationary time series.  Subsequently, we employ both the QPCS and QPCFR procedures for variable screening our previously defined EBIC criterion.  We implement fixed window recursive forecasts with varying window lengths: $l=80,100,120$. This corresponds to forecasting periods of 61, 41, and 21 quarters, respectively, spanning from 2007Q3, 2012Q3, and 2017Q3 to 2022Q3. Table \ref{tab:quarter} provides a summary of the top 5 most frequently selected macroeconomic variables under different forecasting window lengths using both approaches.

	\begin{table}[H]
		\scriptsize\caption{Frequency table of top-5 selected macroeconomic variables by two QPC
			screening procedures, $\tau=0.05$ \label{tab:quarter}}
		
		\begin{centering}
			\begin{tabular}{ccccccccccc}
				\hline 
				& Periods$\left(l\right)$ &  & \multicolumn{2}{c}{2007Q3-2022Q3 $\left(l=80\right)$} &  & \multicolumn{2}{c}{2012Q3-2022Q3 $\left(l=100\right)$} &  & \multicolumn{2}{c}{2017Q3-2022Q3 $\left(l=120\right)$}\tabularnewline
				Method & Rank &  & Variable name & Freq &  & Variable name & Freq &  & Variable name & Freq\tabularnewline
				\cline{1-2} \cline{2-2} \cline{4-5} \cline{5-5} \cline{7-8} \cline{8-8} \cline{10-11} \cline{11-11} 
				& 1 &  & ULCMFG & 0.492  &  & CONSUMERx & 0.390  &  & AWHMAN & 0.524 \tabularnewline
				& 2 &  & TARESAx & 0.410  &  & TARESAx & 0.341  &  & CONSUMERx & 0.524 \tabularnewline
				QPCS & 3 &  & AWHMAN & 0.361  &  & FEDFUNDS & 0.317  &  & CLAIMSx & 0.524 \tabularnewline
				& 4 &  & FEDFUNDS & 0.344  &  & ULCMFG & 0.293  &  & SRVPRD & 0.476 \tabularnewline
				& 5 &  & TFAABSHNOx & 0.311  &  & CLAIMSx & 0.293  &  & CES9091000001 & 0.476 \tabularnewline
				\cline{1-2} \cline{2-2} \cline{4-5} \cline{5-5} \cline{7-8} \cline{8-8} \cline{10-11} \cline{11-11} 
				& 1 &  & ULCMFG & 0.475  &  & CONSUMERx & 0.439  &  & CONSUMERx & 0.524 \tabularnewline
				& 2 &  & TARESAx & 0.443  &  & TARESAx & 0.341  &  & CLAIMSx & 0.524 \tabularnewline
				QPCFR & 3 &  & FEDFUNDS & 0.328  &  & ULCMFG & 0.268  &  & SRVPRD & 0.476 \tabularnewline
				& 4 &  & AWHMAN & 0.279  &  & FEDFUNDS & 0.268  &  & CES9091000001 & 0.429 \tabularnewline
				& 5 &  & CLAIMSx & 0.279  &  & CLAIMSx & 0.244  &  & TARESAx & 0.333 \tabularnewline
				\hline 
			\end{tabular}
			\par\end{centering}
		\begin{note*}
			This table presents the top 5 selected variables obtained from two QPC screening approaches with different fixed window length forecasts $l=80,100,120$. The column labeled
			\textquotedbl Freq\textquotedbl{} denotes the selected frequency
			for each variable. Variable names are documented in \citet{mccracken2020fred}. 
		\end{note*}
	\end{table}
	
	Our findings indicate that both QPCS and QPCFR consistently produce similar results, underscoring the robustness of our analysis. Regardless of the chosen window length, the variable "CLAIMSx," representing the initial claims for unemployment, appears prominently in almost all cases. This suggests the significance of labor market indicators in GaR forecasting. Furthermore, variables such as "CONSUMERx" (real consumer loans at all commercial banks), "FEDFUNDS" (effective federal funds rates), and "TARESAx" (real assets of households and nonprofit organizations excluding real estate assets) are frequently selected by our QPC procedure. These indicators have stronger connections to the financial market or nationwide financial conditions.

	While our fixed rolling window forecasting results provide valuable insights, researchers typically place greater emphasis on GaR during recessionary periods. Therefore, we delve deeper into variable selections during two specific periods: the global financial crisis (2007Q3-2009Q2) and the COVID-19 recession (2020Q1-2022Q3). For the financial crisis period, we utilize fixed rolling window forecasting with a window length of $l=80$ due to data limitations. The top 5 variable selections for both screening methods during this period are presented in Table \ref{tab:financial-crisis}. For the COVID-19 recession period, we apply various window lengths, including $l=80,100,$ and $120$, to ensure the robustness of our findings. The results are summarized in Table \ref{tab:covid19}.
	
	\begin{table}[H]
		\caption{Top-5 selected variables during the global financial crisis, 2007Q3-2009Q2,
			$\tau=0.05$ \label{tab:financial-crisis}}
		\small
		\begin{centering}
			\begin{tabular}{ccccc}
				\hline 
				& $l$ &  & \multicolumn{2}{c}{$80$}\tabularnewline
				Method & Rank &  & Variable name & Times\tabularnewline
				\cline{1-2} \cline{2-2} \cline{4-5} \cline{5-5} 
				& 1 &  & SRVPRD & 5/8\tabularnewline
				& 2 &  & BAA & 5/8\tabularnewline
				QPCS & 3 &  & OILPRICEx & 4/8\tabularnewline
				& 4 &  & NASDAQCOM & 4/8\tabularnewline
				& 5 &  & FPIx & 3/8\tabularnewline
				\cline{1-2} \cline{2-2} \cline{4-5} \cline{5-5} 
				& 1 &  & SRVPRD & 5/8\tabularnewline
				& 2 &  & BAA & 4/8\tabularnewline
				QPCFR & 3 &  & NASDAQCOM & 4/8\tabularnewline
				& 4 &  & FPIx & 3/8\tabularnewline
				& 5 &  & OILPRICEx & 3/8\tabularnewline
				\hline 
			\end{tabular}
			\par\end{centering}
			\footnotesize
		\begin{note*}
			This table presents the top 5 selected variables obtained from two QPC screening approaches during the global financial crisis periods. ``Times'' means
			the number of selected times for each variable out of $8$ periods.  Variable names are documented in \citet{mccracken2020fred}. 
		\end{note*}
	\end{table}
	
	\begin{table}[H]
		\scriptsize
		\caption{Top-5 selected variables during the covid-19 recession, 2020Q1-2022Q3
			$\tau=0.05$ \label{tab:covid19}}
		
		\begin{centering}
			\begin{tabular}{ccccccccccc}
				\hline 
				& $l$ &  & \multicolumn{2}{c}{$80$} &  & \multicolumn{2}{c}{$100$} &  & \multicolumn{2}{c}{$120$}\tabularnewline
				Method & Rank &  & Variable name & Times &  & Variable name & Times &  & Variable name & Times\tabularnewline
				\cline{1-2} \cline{2-2} \cline{4-11} \cline{5-11} \cline{6-11} \cline{7-11} \cline{8-11} \cline{9-11} \cline{10-11} \cline{11-11} 
				& 1 &  & CLAIMSx & 10/11 &  & CLAIMSx & 8/11 &  & CLAIMSx & 11/11\tabularnewline
				& 2 &  & AWHMAN & 9/11 &  & USEHS & 6/11 &  & AWHMAN & 6/11\tabularnewline
				QPCS & 3 &  & LNS13023621 & 6/11 &  & LNS13023621 & 4/11 &  & DPIC96 & 4/11\tabularnewline
				& 4 &  & AHETPIx & 4/11 &  & AWHMAN & 4/11 &  & DHUTRG3Q086SBEA & 4/11\tabularnewline
				& 5 &  & TFAABSHNOx & 4/11 &  & TARESAx & 4/11 &  & TARESAx & 3/11\tabularnewline
				\cline{1-2} \cline{2-2} \cline{4-11} \cline{5-11} \cline{6-11} \cline{7-11} \cline{8-11} \cline{9-11} \cline{10-11} \cline{11-11} 
				& 1 &  & AWHMAN & 8/11 &  & CLAIMSx & 6/11 &  & CLAIMSx & 11/11\tabularnewline
				& 2 &  & CLAIMSx & 8/11 &  & USEHS & 5/11 &  & AWHMAN & 6/11\tabularnewline
				QPCFR & 3 &  & LNS13023621 & 6/11 &  & LNS13023621 & 4/11 &  & TARESAx & 5/11\tabularnewline
				& 4 &  & TFAABSHNOx & 5/11 &  & TARESAx & 4/11 &  & RSAFSx & 3/11\tabularnewline
				& 5 &  & TARESAx & 4/11 &  & AWHMAN & 3/11 &  & TFAABSHNOx & 3/11\tabularnewline
				\hline 
			\end{tabular}
			\par\end{centering}
		\begin{note*}
			This table presents the top 5 selected variables obtained from two QPC screening approaches during the covid-19 recession periods. ``Times'' means
			the number of selected times for each variable out of $11$ periods.  Variable names are documented in \citet{mccracken2020fred}. 
		\end{note*}
	\end{table}
	
	It is evident that the variables selected by our two QPC procedures differ significantly during different recession periods. During the global financial crisis, the selected variables are predominantly indicators related to the financial sector. These include "SRVPRD" (the number of all employees in service-providing industries), "BAA" (Moody's seasoned Baa corporate bond yield), "OILPRICEx" (real crude oil prices), "NASDAQCOM" (NASDAQ composite index), and "FPIx" (the amount of real private fixed investment). In contrast, during the COVID-19 recession period, the selected variables are primarily from the labor market, including "CLAIMSx" (initial claims for unemployment), as previously mentioned, and "AWHMAN" (average weekly hours of production and nonsupervisory employees). These findings suggest a shift in the importance of economic indicators in GaR forecasting during different economic downturns.
	
	Our results also provide valuable insights into GaR prediction that complement existing literature from a different perspective. While much of the previous research focuses on specific financial condition indicators for GaR prediction, such as NFCI in \citet{adrian2019vulnerable} and \citet{brownlees2021backtesting}, our findings suggest the importance of labor market indicators.
	
	In the realm of financial condition variables, our results offer more specific insights than the NFCI, particularly during the global financial crisis. Additionally, our evidence underscores the significance of labor market indicators such as initial claims for unemployment and average working hours, aligning with previous macroeconomics research findings. \citet{schmidt2022climbing} demonstrates that initial claims, serving as a proxy for the labor market risk index, are robust predictors of broad market returns, especially in terms of stock return predictability. Therefore, labor market indicators are poised to play a crucial role in GaR prediction.More precisely, the predictability of the financial condition index on GaR may be partially absorbed by labor market indicators. This intuition is corroborated by our QPC approach, which indicates that labor market indicators are selected more frequently, while fewer financial indicators are chosen.
	
	Another noteworthy outcome of the QPC approach is its ability to assess the predictive power of NFCI, which has been a subject of debate in the existing literature. \citet{adrian2019vulnerable} and \citet{plagborg2020growth} hold opposing views on the effectiveness of NFCI. The former contends that NFCI is a valuable predictor, while the latter argues against its utility. 
	
	To investigate this matter, we incorporate NFCI into our dataset and observe whether it is selected by our QPC approach. If NFCI is selected, it implies that it retains predictive power even when considering other selected predictors as controls. Conversely, if it is not selected, the implication is that NFCI is potentially endogenous and correlated with other macroeconomic variables, indicating the importance of these alternative variables. 
	
	In Table \ref{tab:nfci}, our results reveal that NFCI is only sporadically selected throughout the entire forecasting period, even within a comprehensive dataset of macroeconomic variables. Furthermore, NFCI exhibits no predictive power during the COVID-19 recession, while it contributes only marginally to GaR forecasting during the global financial crisis. This finding contradicts the assertion made by \citet{adrian2019vulnerable}. One plausible explanation is that NFCI is a composite index derived from a weighted average of numerous financial variables. When using a vast dataset and considering variable selection, researchers can identify the most informative predictors, causing NFCI to lose its forecasting efficacy when conditioned on these selected variables.
	
	\begin{table}[H]
		\caption{NFCI selected frequency, $\tau=0.05$ \label{tab:nfci}}
		\small
		\begin{centering}
			\begin{tabular}{ccccccccc}
				\hline 
				& whole periods &  &  &  & \multicolumn{1}{c}{global financial crisis} &  &  & \multicolumn{1}{c}{covid-19 recession}\tabularnewline
				Method & $l$ & Freq &  & $l$ & Freq &  & $l$ & Freq\tabularnewline
				\cline{1-3} \cline{2-3} \cline{3-3} \cline{5-6} \cline{6-6} \cline{8-9} \cline{9-9} 
				& $80$ & 4/61 &  &  &  &  & $80$ & 0/11\tabularnewline
				QPCS & $100$ & 1/41 &  & $80$ & 2/8 &  & $100$ & 0/11\tabularnewline
				& $120$ & 0/21 &  &  &  &  & $120$ & 0/11\tabularnewline
				\cline{1-3} \cline{2-3} \cline{3-3} \cline{5-6} \cline{6-6} \cline{8-9} \cline{9-9} 
				& $80$ & 3/61 &  &  &  &  & $80$ & 0/11\tabularnewline
				QPCFR & $100$ & 1/41 &  & $80$ & 2/8 &  & $100$ & 0/11\tabularnewline
				& $120$ & 0/21 &  &  &  &  & $120$ & 0/11\tabularnewline
				\hline 
			\end{tabular}
			\par\end{centering}
			\footnotesize
		\begin{note*}
			This table presents the selection frequency of NFCI for two QPC screening
			approaches with different window length $l$ during all periods, global
			financial crisis, and covid-19 recession. ``Freq'' means the number
			of selected times for NFCI out of total number of forecasting periods.
		\end{note*}
	\end{table}

	\section{Summary}
	
	In summary, this paper extends the quantile partial correlation approach to variable screening within the context of stationary $\beta$-mixing data. Our theoretical framework provides general results, with the i.i.d. data as a special case. The uniform convergence of QPC and its screening property validate its potential utility. Simulations demonstrate the effectiveness and efficiency of the QPC approach in high-dimensional time-series models. We apply our QPC variable screening method to study GaR forecasting with a large number of predictors, and our novel empirical evidence suggests that labor market indicators hold value in GaR prediction alongside financial market indicators.

	\newpage
	
	\bibliography{qpc}
	
	\appendix
	\section{Proofs}
	
	This appendix section contains supplemental materials, technical proofs of lemmas, theorems, and propositions in the main context.
	
	\subsection{Theorem 1 of \citet{ma2017variable}}
	
	In this subsection, we provide the original version of Theorem 1 in \citet{ma2017variable} given the following two assumptions.
	
	\begin{assumption}\label{assu: C1}[Assumption (C1) in \citet{ma2017variable}]
	 	The conditional density $f_{Y|X=x}\left(y\right)$ of $y$ given $X=x$ satisfies the Lipschitz condition of order 1 and $f_{Y|X=x}\left(y\right)>0$ for any $y$.
	\end{assumption} 

	\begin{assumption}\label{assu: C2}[Assumption (C2) in \citet{ma2017variable}]
	The predictors satisfy the following boundedness conditions for some positive constants $M_1,M_2,M_3$, and $M_4$, such that,  $\sup_{i,j}\left|X_{i,j}\right|\leq M_{1}$, $%
	\sup_{i,j}\left|X_{i,S_{j}}^{T}\theta_{j}^{0}\right|\leq M_{2}$, $%
	\sup_{i,j}\left|X_{i,S_{j}}^{T}\pi_{j}^{0}\right|\leq M_{3}$, and $%
	\sup_{j}\left\Vert \frac{1}{n}\sum_{i=1}^{n}X_{i,S_{j}}X_{i,j}\right\Vert
	\leq M_{4}$. for every $j=1,\dots,p$.
	
    Additionally, for every $1\leq j\leq p$, we assume that 
	\begin{equation*}
		m\leq\lambda_{\min}\left(E\left(X_{S_{j}}X_{S_{j}}^{T}\right)\right)\leq%
		\lambda_{\max}\left(E\left(X_{S_{j}}X_{S_{j}}^{T}\right)\right)\leq M
	\end{equation*}
	for some universal positive constants $m$ and $M$.
\end{assumption} 
	
		\begin{thm*}
		(Theorem 1 in \citet{ma2017variable}) We suppose the data $\{Y_{i+1},X_i\}$ is independent and identically distributed. Under assumptions (C1) and (C2)  in %
		\citet{ma2017variable}, for some universal constant $C$, $0<\kappa <\frac{1}{%
			2}$, and $r_{n}=Cn^{\omega }$ for some $0\leq \omega <\min \left( \left(
		1-2\kappa \right) ,2\kappa \right) $, we have 
		\begin{align*}
			& P\left( \sup_{1\leq j\leq p_{n}}\left\vert \widehat{qpcor}_{\tau }\left(
			Y_{i+1},X_{i,j}|X_{i,S_{j}}\right) -qpcor_{\tau }\left(
			Y_{i+1},X_{i,j}|X_{i,S_{j}}\right) \right\vert \geq Cr_{n}^{\frac{1}{2}%
			}n^{-\kappa }\right) \\
			\leq & p_{n}\left( Cr_{n}^{2}e^{-\frac{Cn^{1-2\kappa }}{r_{n}}%
			}+Cr_{n}^{2}e^{-C\frac{n}{r_{n}^{2}}}\right)
		\end{align*}
	\end{thm*}
	
	There are two minor issues in the proof of this theorem:
	\begin{enumerate}
		\item In the (S.5) inequality and the inequality before (S.5) of %
		\citet{ma2017variable}'s supplemental materials, the correct modifications of the inequalities are 
		\begin{equation*}
			P\left(\left|\lambda_{\min}\left(n^{-1}%
			\sum_{i=1}^{n}X_{i,S_{j}}X_{i,S_{j}}^{T}\right)-\lambda_{\min}%
			\left(EX_{i,S_{j}}X_{i,S_{j}}^{T}\right)\right|\geq
			c_{3}^{*}r_{n}n^{-1}\delta_{1}^{*}\right)\leq2exp\left(-c_{4}^{*}%
			\delta_{1}^{*^{2}}n^{-1}\right)
		\end{equation*}
		and 
		\begin{equation*}
			P\left(\left\Vert
			n^{-1}\sum_{i=1}^{n}X_{i,S_{j}}X_{i,S_{j}}^{T}-E%
			\left(X_{i,S_{j}}X_{i,S_{j}}^{T}\right)\right\Vert \geq
			c_{3}^{*}r_{n}n^{-1}\delta_{1}^{*}\right)\leq2exp\left(-c_{4}^{*}%
			\delta_{1}^{*^{2}}n^{-1}\right)
		\end{equation*}%
		\par
		This result also affects their lemma 2, 4, and 5.
		\par
		\item In the (S.12) inequality of \citet{ma2017variable}'s supplemental
		materials. The Bernstein's inequality would lead to the following result 
		\begin{equation*}
			P\left(\left|\bar{\omega}_{n}\left(\pi_{j}^{0}\right)-\bar{\omega}%
			\left(\pi_{j}^{0}\right)\geq\frac{1}{2}c_{8}^{*}n^{-2\kappa}\right|\right)%
			\leq2exp\left(-c_{12}^{*}n^{1-4\kappa}\right)
		\end{equation*}%
		\par
		This change also affects their lemma 3 and the range of $\kappa$ in theorem 1.
	\end{enumerate}
	After the correction, we have provide the refined theorem of uniform convergence of population QPC to the sample QPC in the main text. It is noteworthy to mention that the refined version is a special case of our theorem \ref{thm:1}.
	\subsection{Introduction of auxiliary lemmas}

	Our goal is to use sample QPC, $\widehat{qpcor_{\tau}}\left(Y_{i+1},X_{i,j}|X_{i,S_{j}}\right)$, from the stationary mixing data process to screen variables.
	Therefore, we establish the
	uniform convergence of $\widehat{qpcor_{\tau}}%
	\left(Y_{i+1},X_{i,j}|X_{i,S_{j}}\right)$ to the population QPC, ${%
		qpcor_{\tau}}\left(Y_{i+1},X_{i,j}|X_{i,S_{j}}\right)$. To show this uniform convergence, we first prove the convergences results of $\hat{\theta}_j$, $\hat{\pi}_j$, $\frac{1}{n}\sum_{i=1}^{n}\psi_{\tau}\left(Y_{i+1}-X_{i,S_{j}}^{T}%
	\hat{\pi}\right)\left(X_{i,j}-X_{i,S_{j}}^{T}\hat{\theta}_{j}\right)$, and $%
	\hat{\sigma}_{j}^{2}$ in the following lemma \ref{lem:2} to lemma %
	\ref{lem:5} and then combine those results to prove our theorems.
	
	\begin{lem}
		\label{lem:2} Under assumption \ref{assu:beta-mix} and \ref{assu:bound}, by assuming $%
		n^{-1}\delta_{n}=o\left(1\right)$ and $n^{-1}\delta_{1}=o\left(1\right)$,
		denote 
		\begin{equation*}
			\hat{\theta}_{j}=\left(\frac{1}{n}\sum_{i=1}^{n}X_{i,S_{j}}X_{i,S_{j}}^{T}%
			\right)^{-1}\left(\frac{1}{n}\sum_{i=1}^{n}X_{i,S_{j}}X_{i,j}\right)
		\end{equation*}
		and 
		\begin{equation*}
			\theta_{j}^{0}=E\left(X_{i,S_{j}}X_{i,S_{j}}^{T}\right)^{-1}E%
			\left(X_{i,S_{j}}X_{i,j}\right)
		\end{equation*}
		as the sample estimated coefficient and population coefficient of $%
		X_{i,S_{j}}$ on $X_{i,j}$. We have 
		\begin{equation*}
			P\left(\left\Vert \hat{\theta}_{j}-\theta_{j}^{0}\right\Vert \geq\frac{%
				C_{1}r_{n}^{\frac{1}{2}}\delta_{n}}{mn}+\frac{C_{2}r_{n}\delta_{1}}{n}%
			\right)\leq r_{n}e^{-\frac{C_{3}\delta_{n}^{2}}{n}}+e^{-\frac{C_{5}n}{%
					r_{n}^{2}}}+e^{-\frac{C_{4}\delta_{1}^{2}}{n}}
		\end{equation*}
		
		Moreover, if $\delta_{1}=\delta_{n}$, we have
		
		\begin{equation*}
			P\left(\left\Vert \hat{\theta}_{j}-\theta_{j}^{0}\right\Vert \geq C_{6}\frac{%
				r_{n}\delta_{n}}{n}\right)\leq C_{7}r_{n}e^{-\frac{C_{3}\delta_{n}^{2}}{n}%
			}+e^{-\frac{C_{5}n}{r_{n}^{2}}}
		\end{equation*}
	\end{lem}
	
	\begin{lem}
		\label{lem:3} Under assumption \ref{assu:beta-mix}, \ref{assu:bound}, and \ref{assu:eigen},
		for any $1\leq j\leq p_{n}$ and the mixing coefficient $%
		B$, we have for some $0<\kappa<\frac{1}{4}$, 
		\begin{equation*}
			P\left(\left\Vert \hat{\pi}_{j}-\pi_{j}^{0}\right\Vert \geq
			C_{8}n^{-\kappa}\right)   \leq e^{-C_{9}n^{1-4\kappa}}+C_{10}e^{-C_{11}\left(\frac{n^{1-2\kappa}}{r_{n}}\right)}+C_{12}n^{\frac{1}{2}}e^{-Bn^{\frac{1}{2}}}
		\end{equation*}
		where $\hat{\pi}_{j}=\arg\min\bar{\omega}_{n}\left(\pi_{j}\right)=\arg\min%
		\frac{1}{n}\sum_{i=1}^{n}\left(\rho_{\tau}\left(Y_{i+1}-X_{i,S_{j}}^{T}%
		\pi_{j}\right)-\rho_{\tau}\left(Y_{i+1}\right)\right)$ and $%
		\pi_{j}^{0}=\arg\min\bar{\omega}\left(\pi_{j}\right)=\arg\min
		E\left(\rho_{\tau}\left(Y_{i+1}-X_{i,S_{j}}^{T}\pi_{j}\right)-\rho_{\tau}%
		\left(Y_{i+1}\right)\right)$.
	\end{lem}

	\begin{lem}
		\label{lem:4} Under assumption \ref{assu:beta-mix}, \ref{assu:bound}, and \ref{assu:eigen},
		for any $1\leq j\leq p_{n}$ and the mixing coefficient $B$, we have for some $0<\kappa<\frac{1}{4}$, 
		\begin{align*}
			& P\left\{ \left|\frac{1}{n}\sum_{i=1}^{n}\psi_{\tau}%
			\left(Y_{i+1}-X_{i,S_{j}}^{T}\hat{\pi}_{j}\right)%
			\left(X_{i,j}-X_{i,S_{j}}^{T}\hat{\theta}_{j}\right)-E\left[%
			\psi_{\tau}\left(Y_{i+1}-X_{i,S_{j}}^{T}\pi_{j}^{0}\right)X_{i,j}\right]%
			\right|\geq C_{13}r_{n}^{\frac{1}{2}}n^{-\kappa}\right\} \\
			&
			C_{14}e^{-C_{15}n^{1-4\kappa}}+C_{10}e^{-C_{11}r_{n}^{-1}n^{1-2\kappa}}+C_{12}n^{%
				\frac{1}{2}}e^{-Bn^{\frac{1}{2}}}+C_{16}r_{n}e^{-C_{17}\frac{n}{%
					r_{n}^{2}}}
		\end{align*}
	\end{lem}
	
	\begin{lem}
		\label{lem:5} Under assumptions \ref{assu:beta-mix}, \ref{assu:bound}, and \ref{assu:eigen}%
		, for any $1\leq j\leq p_{n}$, there exists some universal
		constants $C_i$s such that 
		\begin{equation*}
			P\left(\left|\hat{\sigma}_{j}^{2}-\sigma_{j}^{2}\right|\geq C_{18}r_{n}^{%
				\frac{1}{2}}n^{-\kappa}\right)\leq e^{-\frac{C_{19}n}{r_{n}}}+C_{20}e^{-%
				\frac{C_{21}n}{r_{n}^{2}}}+C_{22}r_{n}e^{-\frac{C_{23}n^{1-2\kappa}}{r_{n}}}
		\end{equation*}
		
		In addition, with assumption \ref{assu:density}, we have for $a\in\left(0,1\right)$%
		, 
		\begin{equation*}
			P\left(\left|\hat{\sigma}_{j}^{2}-\sigma_{j}^{2}\right|\geq
			a\sigma_{j}^{2}\right)\leq e^{-\frac{C_{19}n}{r_{n}}}+C_{20}e^{-\frac{C_{21}n%
				}{r_{n}^{2}}}+C_{22}r_{n}e^{-\frac{C_{23}n^{1-2\kappa}}{r_{n}}}
		\end{equation*}
	\end{lem}

	\subsection{Proof of lemma \protect\ref{lem:2}}
	
	\begin{proof} We can represent $\hat{\theta}_{j}-\theta_{j}^{0}$ as 
		\begin{align*}
			\hat{\theta}_{j}-\theta_{j}^{0}= &
			E\left(X_{i,S_{j}}X_{i,S_{j}}^{T}\right)^{-1}\left(\frac{1}{n}%
			\sum_{i=1}^{n}X_{i,S_{j}}X_{i,j}^{T}-E\left(X_{i,S_{j}}X_{i,j}\right)\right)
			\\
			+ & \left(\left(\frac{1}{n}\sum_{i=1}^{n}X_{i,S_{j}}X_{i,S_{j}}^{T}%
			\right)^{-1}-E\left(X_{i,S_{j}}X_{i,S_{j}}^{T}\right)^{-1}\right)\left(\frac{%
				1}{n}\sum_{i=1}^{n}X_{i,S_{j}}X_{i,j}\right) \\
			:= & \Gamma_{1j}+\Gamma_{2j}
		\end{align*}
		
		Now, consider $T_{ikj}=X_{i,k}X_{i,j}-E\left(X_{i,k}X_{i,j}\right)$ for $%
		k\in\left\{ 0\right\} \cup S_{j}$. Under assumption \ref{assu:beta-mix}, we have $%
		\left\{ T_{ijk}\right\} _{i=1}^{n}$ is $\alpha$-mixing with $%
		\alpha\left(n\right)\leq e^{-Bn}$. Using assumption \ref{assu:bound}, we
		have $\left|T_{ijk}\right|\leq2M_{1}^{2}$. Use Theorem 1 in %
		\citet{merlevede2009bernstein}, we have, 
		\begin{align*}
			P\left(\frac{1}{n}\left|\sum_{i=1}^{n}T_{ikj}\right|\geq\frac{c_{1}\delta_{n}%
			}{n}\right) & \leq e^{-\frac{cc_{1}^{2}\delta_{n}^{2}}{n(2M_{1}^{2})^{2}+%
					\left(2M_{1}^{2}\right)c_{1}\delta_{n}\left(\log n\right)\left(\log\log
					n\right)}} \\
			& \leq e^{-\frac{c_{2}\delta_{n}^{2}}{n}}
		\end{align*}
		for sufficiently large $n$. The last inequality is based on $%
		n^{-1}\delta_{n}=o\left(1\right)$.\footnote{\begin{rem*}
				If $\frac{\delta_{n}}{n}=O\left(1\right)$, we can only obtain $P\left(\frac{1%
				}{n}\left|\sum_{i=1}^{n}T_{ikj}\right|\geq\frac{c_{1}\delta_{n}}{n}%
				\right)\leq e^{-\frac{c {\delta}_{n}^{2}}{n\log n\log\log n}}$. Here, we assume 
				$r_{n}=Cn^{\gamma}$ for some constant $\gamma$, then $\frac{\delta_{n}}{n}%
				=o\left(1\right)$ is satisfied in lemma \ref{lem:4}.
			\end{rem*}
		}
		
		Since $r_{n}=\max_{1\leq j\leq p}\left|S_{j}\right|$, using the Bonferroni
		bound, we have 
		\begin{equation*}
			P\left(\left\Vert \frac{1}{n}\sum_{i=1}^{n}X_{i,S_{j}}X_{i,j}-E%
			\left(X_{i,S_{j}}X_{i,j}\right)\right\Vert \geq\frac{c_{1}r_{n}^{\frac{1}{2}%
				}\delta_{n}}{n}\right)\leq r_{n}e^{-\frac{c_{2}\delta_{n}^{2}}{n}}
		\end{equation*}
		
		Using the assumption \ref{assu:eigen}, we obtain 
		\begin{equation*}
			P\left(\left\Vert \Gamma_{1j}\right\Vert \geq\frac{c_{1}r_{n}^{\frac{1}{2}%
				}\delta_{n}}{mn}\right)\leq r_{n}e^{-\frac{c_{2}\delta_{n}^{2}}{n}}
		\end{equation*}
		
		Now we consider $D_{j}:=\frac{1}{n}%
		\sum_{i=1}^{n}X_{i,S_{j}}X_{i,S_{j}}^{T}-E\left(X_{i,S_{j}}X_{i,S_{j}}^{T}%
		\right)$ and $D_{j,kk^{^{\prime }}}=\frac{1}{n}%
		\sum_{i=1}^{n}X_{i,k}X_{i,k^{^{\prime }}}-E\left(X_{i,k}X_{i,k^{^{\prime
		}}}\right)$ for $k,k^{^{\prime }}\in\left\{ 0\right\} \cup S_{j}$. Hence,
		under assumption \ref{assu:beta-mix}, we have $\left\{ D_{j,kk^{^{\prime
		}}}\right\} _{j}$ is $\alpha$-mixing with $\alpha\left(n\right)\leq
		e^{-Bn}$. Moreover, we have $\left|X_{i,k}X_{i,k^{^{\prime
		}}}\right|\leq2M_{1}^{2}$ and $var\left(X_{i,k}X_{i,k^{^{\prime
		}}}\right)\leq M_{1}^{4}$. Use Theorem 1 in \citet{merlevede2009bernstein},
		we have, 
		
		\begin{align*}
			P\left(\left|D_{j,kk^{^{\prime }}}\right|\geq\frac{c_{3}\delta_{n}}{n}%
			\right) & \leq e^{-\frac{cc_{3}^{2}\delta_{n}^{2}}{n(2M_{1}^{2})^{2}+%
					\left(2M_{1}^{2}\right)c_{3}\delta_{n}\left(\log n\right)\left(\log\log
					n\right)}} \\
			& \leq e^{-\frac{c_{4}\delta_{n}^{2}}{n}}
		\end{align*}
		for suffiently large $n$ and all $\delta_{n}$, $k$, $k^{^{\prime }}$. The last
		inequality is from $n^{-1}\delta_{n}=o\left(1\right)$.
		
		In addition, we have 
		\begin{align*}
			\left\Vert \frac{1}{n}\sum_{i=1}^{n}X_{i,S_{j}}X_{i,S_{j}}^{T}-E%
			\left(X_{i,S_{j}}X_{i,S_{j}}^{T}\right)\right\Vert & =\lambda_{\max}\left(%
			\frac{1}{n}\sum_{i=1}^{n}X_{i,S_{j}}X_{i,S_{j}}^{T}-E%
			\left(X_{i,S_{j}}X_{i,S_{j}}^{T}\right)\right) \\
			& \leq\sum_{k\in\left\{ 0\right\} \cup S_{j}}\left(\frac{1}{n}%
			\sum_{i=1}^{n}X_{i,kk}X_{i,kk}-E\left(X_{i,k}X_{i,k}\right)\right) \\
			& \leq r_{n}\max_{k}\left|D_{j,kk}\right|
		\end{align*}
		and, using Weyl's inequality, we can show that 
		\begin{align*}
			\left|\lambda_{\min}\left(\frac{1}{n}%
			\sum_{i=1}^{n}X_{i,S_{j}}X_{i,S_{j}}^{T}\right)+\lambda_{\min}\left(-E%
			\left(X_{i,S_{j}}X_{i,S_{j}}^{T}\right)\right)\right| &
			\leq\left|\lambda_{\min}\left(\frac{1}{n}%
			\sum_{i=1}^{n}X_{i,S_{j}}X_{i,S_{j}}^{T}-E\left(X_{i,S_{j}}X_{i,S_{j}}^{T}%
			\right)\right)\right| \\
			& \leq\left|\left\Vert \frac{1}{n}\sum_{i=1}^{n}X_{i,S_{j}}X_{i,S_{j}}^{T}-E%
			\left(X_{i,S_{j}}X_{i,S_{j}}^{T}\right)\right\Vert \right| \\
			& \leq r_{n}\max_{k}\left|D_{j,kk}\right|
		\end{align*}
		
		Therefore, we obtain 
		\begin{equation*}
			P\left(\left|\lambda_{\min}\left(\frac{1}{n}%
			\sum_{i=1}^{n}X_{i,S_{j}}X_{i,S_{j}}^{T}\right)+\lambda_{\min}\left(-E%
			\left(X_{i,S_{j}}X_{i,S_{j}}^{T}\right)\right)\right|\geq r_{n}\frac{%
				c_{3}\delta_{1}^{*}}{n}\right)\leq e^{-\frac{c_{4}\delta_{1}^{*^{2}}}{n}}
		\end{equation*}
		and 
		\begin{equation*}
			P\left(\left\Vert \frac{1}{n}\sum_{i=1}^{n}X_{i,S_{j}}X_{i,S_{j}}^{T}-E%
			\left(X_{i,S_{j}}X_{i,S_{j}}^{T}\right)\right\Vert \geq r_{n}\frac{%
				c_{3}\delta_{1}}{n}\right)\leq e^{-\frac{c_{4}\delta_{1}^{^{2}}}{n}}
		\end{equation*}
		for some $\delta_{1}^{*}$ and $\delta_{1}$. Now we consider $\delta_{1}^{*}=\frac{%
			c_{5}nm}{c_{3}r_{n}}$ for some positive constant $c_{5}$ and together with
		assumption \ref{assu:eigen}, we show that 
		\begin{equation*}
			P\left(\left|\lambda_{\min}\left(\frac{1}{n}%
			\sum_{i=1}^{n}X_{i,S_{j}}X_{i,S_{j}}^{T}\right)-\lambda_{\min}\left(E%
			\left(X_{i,S_{j}}X_{i,S_{j}}^{T}\right)\right)\right|\geq
			c_{5}\lambda_{\min}\left(E\left(X_{i,S_{j}}X_{i,S_{j}}^{T}\right)\right)%
			\right)\leq e^{-\frac{c_{4}c_{5}^{2}m^{2}n}{c_{3}^{2}r_{n}^{2}}}
		\end{equation*}
		
		Using the fact that $\left|a^{-1}-b^{-1}\right|\geq
		k^{*}b^{-1}\implies\left|a-b\right|\geq kb$ with $k^{*}=\frac{1}{1-k}-1$ for
		any $k\in\left(0,1\right)$, we have 
		\begin{align*}
			P\left(\left|\lambda_{\min}^{-1}\left(\frac{1}{n}%
			\sum_{i=1}^{n}X_{i,S_{j}}X_{i,S_{j}}^{T}\right)-\lambda_{\min}^{-1}\left(E%
			\left(X_{i,S_{j}}X_{i,S_{j}}^{T}\right)\right)\right|\geq\left(\frac{1}{%
				1-c_{5}}-1\right)\lambda_{\min}^{-1}\left(E\left(X_{i,S_{j}}X_{i,S_{j}}^{T}%
			\right)\right)\right) \\
			\leq e^{-\frac{c_{4}c_{5}^{2}m^{2}n}{c_{3}^{2}r_{n}^{2}}}
		\end{align*}
		which indicates 
		\begin{align*}
			P\left(\lambda_{\min}^{-1}\left(\frac{1}{n}%
			\sum_{i=1}^{n}X_{i,S_{j}}X_{i,S_{j}}^{T}\right)\geq\left(\frac{1}{1-c_{5}}%
			\right)\lambda_{\min}^{-1}\left(E\left(X_{i,S_{j}}X_{i,S_{j}}^{T}\right)%
			\right)\right) & \leq e^{-\frac{c_{4}c_{5}^{2}m^{2}n}{c_{3}^{2}r_{n}^{2}}} \\
			P\left(\lambda_{\min}^{-1}\left(\frac{1}{n}%
			\sum_{i=1}^{n}X_{i,S_{j}}X_{i,S_{j}}^{T}\right)\geq\left(\frac{1}{1-c_{5}}%
			\right)m^{-1}\right) & \leq e^{-\frac{c_{4}c_{5}^{2}m^{3}n}{%
					c_{1}^{2}r_{n}^{2}}} \\
			P\left(\left\Vert \left(\frac{1}{n}\sum_{i=1}^{n}X_{i,S_{j}}X_{i,S_{j}}^{T}%
			\right)^{-1}\right\Vert \geq\left(\frac{1}{1-c_{5}}\right)m^{-1}\right) &
			\leq e^{-\frac{c_{4}c_{5}^{2}m^{2}n}{c_{3}^{2}r_{n}^{2}}} \\
			P\left(\left\Vert \left(\frac{1}{n}\sum_{i=1}^{n}X_{i,S_{j}}X_{i,S_{j}}^{T}%
			\right)^{-1}\right\Vert \left\Vert
			E\left(X_{i,S_{j}}X_{i,S_{j}}^{T}\right)^{-1}\right\Vert \geq\left(\frac{1}{%
				1-c_{5}}\right)m^{-2}\right) & \leq e^{-\frac{c_{4}c_{5}^{2}m^{2}n}{%
					c_{3}^{2}r_{n}^{2}}}
		\end{align*}
		
		Therefore, since by definition 
		\begin{equation*}
			\left\Vert \Gamma_{2j}\right\Vert =\left\Vert \left(\frac{1}{n}%
			\sum_{i=1}^{n}X_{i,S_{j}}X_{i,S_{j}}^{T}\right)^{-1}E%
			\left(X_{i,S_{j}}X_{i,S_{j}}^{T}\right)^{-1}\left(E%
			\left(X_{i,S_{j}}X_{i,S_{j}}^{T}\right)-\frac{1}{n}%
			\sum_{i=1}^{n}X_{i,S_{j}}X_{i,S_{j}}^{T}\right)\left(\frac{1}{n}%
			\sum_{i=1}^{n}X_{i,S_{j}}X_{i,j}\right)\right\Vert
		\end{equation*}
		we can show 
		\begin{align*}
			\left\Vert \Gamma_{2j}\right\Vert & \leq\left\Vert \left(\frac{1}{n}%
			\sum_{i=1}^{n}X_{i,S_{j}}X_{i,S_{j}}^{T}\right)^{-1}\right\Vert \left\Vert
			E\left(X_{i,S_{j}}X_{i,S_{j}}^{T}\right)^{-1}\right\Vert \\
			& \times\left\Vert \frac{1}{n}\sum_{i=1}^{n}X_{i,S_{j}}X_{i,S_{j}}^{T}-E%
			\left(X_{i,S_{j}}X_{i,S_{j}}^{T}\right)\right\Vert \left\Vert \frac{1}{n}%
			\sum_{i=1}^{n}X_{i,S_{j}}X_{i,j}\right\Vert
		\end{align*}
		
		Hence, using assumption \ref{assu:bound} and \ref{assu:eigen}, we have 
		\begin{align*}
			P\left(\left\Vert \Gamma_{2j}\right\Vert \geq\left(\frac{1}{1-c_{5}}%
			\right)m^{-2}r_{n}\frac{c_{3}\delta_{1}}{n}M_{4}\right) & \leq e^{-\frac{%
					c_{4}c_{5}^{2}m^{2}n}{c_{3}^{2}r_{n}^{2}}}+e^{-\frac{c_{4}\delta_{1}^{2}}{n}}
			\\
			P\left(\left\Vert \Gamma_{2j}\right\Vert \geq\frac{c_{6}r_{n}\delta_{1}}{n}%
			\right) & \leq e^{-\frac{c_{7}n}{r_{n}^{2}}}+e^{-\frac{c_{4}\delta_{1}^{2}}{n%
			}}
		\end{align*}
		where $c_{6}=\left(\frac{1}{1-c_{5}}\right)m^{-2}c_{3}M_{4}$ and $c_{7}=\frac{c_{4}c_{5}^{2}m^{2}}{c_{3}^{2}}$.
		
		Hence, by the triangular inequality and Bonferroni bound, we have
		
		\begin{equation*}
			P\left(\left\Vert \hat{\theta}_{j}-\theta_{j}^{0}\right\Vert \geq\frac{%
				C_{1}r_{n}^{\frac{1}{2}}\delta_{n}}{mn}+\frac{C_{2}r_{n}\delta_{1}}{n}%
			\right)\leq r_{n}e^{-\frac{C_{3}\delta_{n}^{2}}{n}}+e^{-\frac{C_{5}n}{%
					r_{n}^{2}}}+e^{-\frac{C_{4}\delta_{1}^{2}}{n}}
		\end{equation*}
		where $C_1 = c_1$, $C_2 = c_6$, $C_3 = c_2$, $C_4 = c_4$, and $C_5=c_7$.
		
		Thus, we have shown the probabilistic bound for $\left\Vert \hat{\theta}%
		_{j}-\theta_{j}^{0}\right\Vert $. If $\delta_{1}=\delta_{n}$, we can directly obtain 
		\begin{equation*}
			P\left(\left\Vert \hat{\theta}_{j}-\theta_{j}^{0}\right\Vert \geq C_{6}\frac{%
				r_{n}\delta_{n}}{n}\right)\leq C_{7}r_{n}e^{-\frac{C_{3}\delta_{n}^{2}}{n}%
			}+e^{-\frac{C_{5}n}{r_{n}^{2}}}
		\end{equation*}
		for some positive constant $C_{6}$ and $C_{7}$. 
	\end{proof}
	
	\subsection{Proof of lemma \protect\ref{lem:3}}
	
	\begin{proof} Based on the inequality (S.9) in \citet{ma2017variable}, we
		have 
		\begin{equation*}
			P\left(\left\Vert \hat{\pi}_{j}-\pi_{j}^{0}\right\Vert \geq
			C_{8}n^{-\kappa}\right)\leq P\left(\sup_{\left\Vert
				\pi_{j}-\pi_{j}^{0}\right\Vert \leq C_{8}n^{-\kappa}}\left|\bar{\omega}%
			_{n}\left(\pi_{j}\right)-\bar{\omega}\left(\pi_{j}\right)\right|\geq\frac{1}{%
				2}\inf_{\left\Vert \pi_{j}-\pi_{j}^{0}\right\Vert =C_{8}n^{-\kappa}}\bar{%
				\omega}\left(\pi_{j}\right)-\bar{\omega}\left(\pi_{j}^{0}\right)\right)
		\end{equation*}
		for some positive constant $C_8$.
		
		Consider the case that $\pi_{j}=\pi_{j}^{0}+C_{8}n^{-\kappa}u$ with some $u$
		satisfying $\left\Vert u\right\Vert =1$. By the Knight's identity from %
		\citet{knight1998limiting}, we show 
		\begin{align*}
			& \bar{\omega}\left(\pi_{j}\right)-\bar{\omega}\left(\pi_{j}^{0}\right) \\
			= &
			E\left(\rho_{\tau}\left(Y_{i+1}-X_{i,S_{j}}^{T}%
			\pi_{j}^{0}-X_{i,S_{j}}^{T}C_{8}n^{-\kappa}u\right)-\rho_{\tau}%
			\left(Y_{i+1}-X_{i,S_{j}}^{T}\pi_{j}^{0}\right)\right) \\
			= &
			E\left(-C_{8}n^{-\kappa}X_{i,S_{j}}^{T}u\left(\tau-1%
			\left(Y_{i+1}-X_{i,S_{j}}^{T}\pi_{j}^{0}\leq0\right)\right)\right) \\
			&
			+E\left(\int_{0}^{X_{i,S_{j}}^{T}C_{8}n^{-\kappa}u}1%
			\left(Y_{i+1}-X_{i,S_{j}}^{T}\pi_{j}^{0}\leq
			s\right)-1\left(Y_{i+1}-X_{i,S_{j}}^{T}\pi_{j}^{0}\leq0\right)ds\right) \\
			= &
			E_{X}\left(\int_{0}^{X_{i,S_{j}}^{T}C_{8}n^{-\kappa}u}f_{Y_{i+1}|X_i}\left(%
			\zeta\right)sds\right)
		\end{align*}
		where $\zeta\in\left(X_{i,S_{j}}^{T}\pi_{j}^{0},X_{i,S_{j}}^{T}\pi_{j}^{0}+s%
		\right)$.
		
		Using assumption \ref{assu:beta-mix}, \ref{assu:bound}, and \ref{assu:eigen} with $%
		\left\Vert u\right\Vert =1$, we have 
		\begin{equation*}
			\bar{\omega}\left(\pi_{j}\right)-\bar{\omega}\left(\pi_{j}^{0}%
			\right)=c_{8}E\left(\left(X_{i,S_{j}}^{T}C_{8}n^{-\kappa}u\right)^{2}%
			\right)\geq c_{8}C_{8}mn^{-2\kappa}
		\end{equation*}
		for some positive constant $c_{8}$.
		
		Hence, we show that for some positive constant $c_{9}$, 
		\begin{equation*}
			\inf_{\left\Vert \pi_{j}-\pi_{j}^{0}\right\Vert =C_{8}n^{-\kappa}}\bar{\omega%
			}\left(\pi_{j}\right)-\bar{\omega}\left(\pi_{j}^{0}\right)\geq
			c_{9}n^{-2\kappa}
		\end{equation*}
		
		Therefore, by the triangular inequality, we know 
		\begin{align*}
			P\left(\left\Vert \hat{\pi}_{j}-\pi_{j}^{0}\right\Vert \geq
			C_{8}n^{-\kappa}\right) & \leq P\left(\sup_{\left\Vert
				\pi_{j}-\pi_{j}^{0}\right\Vert \leq C_{8}n^{-\kappa}}\left|\bar{\omega}%
			_{n}\left(\pi_{j}\right)-\bar{\omega}\left(\pi_{j}\right)\right|\geq\frac{1}{%
				2}\inf_{\left\Vert \hat{\pi}_{j}-\pi_{j}^{0}\right\Vert =C_{8}n^{-\kappa}}%
			\bar{\omega}\left(\pi_{j}\right)-\bar{\omega}\left(\pi_{j}^{0}\right)\right)
			\\
			& \leq P\left(\sup_{\left\Vert \pi_{j}-\pi_{j}^{0}\right\Vert \leq
				C_{8}n^{-\kappa}}\left|\bar{\omega}_{n}\left(\pi_{j}\right)-\bar{\omega}%
			\left(\pi_{j}\right)\right|\geq\frac{1}{2}c_{9}n^{-2\kappa}\right) \\
			& \leq P\left(\left|\bar{\omega}_{n}\left(\pi_{j}^{0}\right)-\bar{\omega}%
			\left(\pi_{j}^{0}\right)\right|\geq\frac{1}{2}c_{9}n^{-2\kappa}\right) \\
			& +\left(\sup_{\left\Vert \pi_{j}-\pi_{j}^{0}\right\Vert \leq
				C_{8}n^{-\kappa}}\left|\bar{\omega}_{n}\left(\pi_{j}\right)-\bar{\omega}%
			_{n}\left(\pi_{j}^{0}\right)-\bar{\omega}\left(\pi_{j}\right)+\bar{\omega}%
			\left(\pi_{j}^{0}\right)\right|\geq\frac{1}{2}c_{9}n^{-2\kappa}\right) \\
			& :=\Gamma_{3}+\Gamma_{4}
		\end{align*}
		
		For $\Gamma_{3}$, by \citet{knight1998limiting}'s identity and assumption %
		\ref{assu:bound}, we have 
		\begin{align*}
			\rho_{\tau}\left(Y_{i+1}-X_{i,S_{j}}^{T}\pi_{j}^{0}\right)-\rho_{\tau}%
			\left(Y_{i+1}\right) &
			=-X_{i,S_{j}}^{T}\pi_{j}^{0}\left(\tau-1\left(Y_{i+1}\leq0\right)\right)+%
			\int_{0}^{X_{i,S_{j}}^{T}\pi_{j}^{0}}1\left(Y_{i+1}\leq
			s\right)-1\left(Y_{i+1}\leq0\right)ds \\
			& \leq c_{10}\sup_{i,j}\left|X_{i,S_{j}}^{T}\pi_{j}^{0}\right| \\
			& \leq c_{10}M_{3}
		\end{align*}
		
		Since $\rho_{\tau}\left(Y_{i+1}-X_{i,S_{j}}^{T}\pi_{j}^{0}\right)-\rho_{%
			\tau}\left(Y_{i+1}\right)$ is an $\alpha$-mixing sequence, using theorem 1
		in \citet{merlevede2009bernstein}, we obtain 
		\begin{align*}
			\Gamma_{3} & =P\left(\left|\bar{\omega}_{n}\left(\pi_{j}^{0}\right)-\bar{%
				\omega}\left(\pi_{j}^{0}\right)\right|\geq\frac{1}{2}c_{9}n^{-2\kappa}\right)
			\\
			&
			=P\left(\left|\sum_{i=1}^{n}\rho_{\tau}\left(Y_{i+1}-X_{i,S_{j}}^{T}%
			\pi_{j}^{0}\right)-\rho_{\tau}\left(Y_{i+1}\right)-E\left(\rho_{\tau}%
			\left(Y_{i+1}-X_{i,S_{j}}^{T}\pi_{j}^{0}\right)-\rho_{\tau}\left(Y_{i+1}%
			\right)\right)\right|\geq\frac{1}{2}c_{9}n^{1-2\kappa}\right) \\
			& \leq e^{-\frac{\frac{1}{4}c c_{9}^{2}n^{2-4\kappa}}{%
					nc_{10}^{2}M_{3}^{2}+c_{10}M_{3}\frac{1}{2}c_{9}n^{1-2\kappa}\log n\log\log n}} \\
			& \leq e^{-c_{11}n^{1-4\kappa}}
		\end{align*}
		
		For $\Gamma_{4}$, we consider $V_{ij}\left(\pi_{j}\right)=\rho_{\tau}%
		\left(Y_{i+1}-X_{i,S_{j}}^{T}\pi_{j}\right)-\rho_{\tau}%
		\left(Y_{i+1}-X_{i,S_{j}}^{T}\pi_{j}^{0}\right)$. Using %
		\citet{knight1998limiting}'s identity again, we show
		
		\begin{align*}
			V_{ij}\left(\pi_{j}\right) &
			=-\left(X_{i,S_{j}}^{T}\pi_{j}-X_{i,S_{j}}^{T}\pi_{j}^{0}\right)\left(\tau-1%
			\left(Y_{i+1}-X_{i,S_{j}}^{T}\pi_{j}^{0}\leq0\right)\right) \\
			&
			+\int_{0}^{X_{i,S_{j}}^{T}\pi_{j}-X_{i,S_{j}}^{T}\pi_{j}^{0}}1%
			\left(Y_{i+1}-X_{i,S_{j}}^{T}\pi_{j}^{0}\leq
			s\right)-1\left(Y_{i+1}-X_{i,S_{j}}^{T}\pi_{j}^{0}\leq0\right)ds
		\end{align*}
		
		Therefore, by assumption \ref{assu:bound}, we have 
		\begin{equation*}
			\sup_{\left\Vert \pi_{j}-\pi_{j}^{0}\right\Vert \leq
				C_{8}n^{-\kappa}}\left|V_{ij}\left(\pi_{j}\right)\right|\leq\sup_{\left\Vert
				\pi_{j}-\pi_{j}^{0}\right\Vert \leq
				C_{8}n^{-\kappa}}2\left|X_{i,S_{j}}^{T}\pi_{j}-X_{i,S_{j}}^{T}\pi_{j}^{0}%
			\right|\leq2\sup_{i,j}\left\Vert X_{i,S_{j}}\right\Vert \left\Vert
			\pi_{j}-\pi_{j}^{0}\right\Vert \leq2r_{n}^{\frac{1}{2}}M_{1}C_{8}n^{-\kappa}
		\end{equation*}
		
		Then we can represent 
		\begin{equation*}
			\Gamma_{4}=P\left(\sup_{\left\Vert \pi_{j}-\pi_{j}^{0}\right\Vert \leq
				C_{8}n^{-\kappa}}\left|\frac{1}{n}\sum_{i=1}^{n}V_{ij}\left(\pi_{j}%
			\right)-EV_{ij}\left(\pi_{j}\right)\right|\geq\frac{1}{2}c_{3}n^{-2\kappa}%
			\right)
		\end{equation*}

		To obtain the upper bound of $\Gamma_{4}$, we rely on the blocking technique from \citet{yu1994rates} and \citet{arcones1994central} and the Talagrand's concentration inequality of $\beta$-mixing
		process from \citet{galvao2016smoothed}.
		
		The basic idea of blocking technique allows us to construct our $n$
		samples into blocks with $q\in\left\lfloor 1,\frac{n}{2}\right\rfloor $.
		We define the blocks in the following way
		\begin{align*}
			H_{k}= & \left\{ i:2\left(k-1\right)q+1\leq i\leq\left(2k-1\right)q\right\} \\
			T_{k}= & \left\{ i:\left(2k-1\right)q+1\leq i\leq2kq\right\} 
		\end{align*}
		, where $k=1,\dots,r$. Therefore, we have $r=\left\lfloor \frac{n}{2q}\right\rfloor $
		number of $H_{k}$ and $T_{k}$ for each. 
		
		Now, since we have $V_{ij}\left(\pi_{j}\right)-EV_{ij}\left(\pi_{j}\right)$
		has mean $0$ and $\sup_{\left\Vert \pi_{j}-\pi_{j}^{0}\right\Vert \leq C_{8}n^{-\kappa}}\left|V_{ij}\left(\pi_{ij}\right)\right|\leq2r_{n}^{\frac{1}{2}}M_{1}C_{8}n^{-\kappa}$,
		by the corollary C.1. in \citet{galvao2016smoothed}, we have 
		\begin{align*}
			P & \Biggl\{\sup_{\left\Vert \pi_{j}-\pi_{j}^{0}\right\Vert \leq C_{8}n^{-\kappa}}\left|\frac{1}{n}\sum_{i=1}^{n}V_{ij}\left(\pi_{ij}\right)-EV_{ij}\left(\pi_{j}\right)\right|\\
			& \geq C^{*}\left[\frac{\sqrt{\left(s\vee1\right)n}}{n}\sigma_{q}\left(V_{ij}\left(\pi_{j}\right)-EV_{ij}\left(\pi_{j}\right)\right)+\frac{sq}{n}2r_{n}^{\frac{1}{2}}M_{1}C_{8}n^{-\kappa}\right]\Biggl\}\\
			\leq & 2e^{-s}+2r\beta\left(q\right)
		\end{align*}
		where $\sigma_{q}\left(V_{ij}\left(\pi_{j}\right)-EV_{ij}\left(\pi_{j}\right)\right)$
		is the long-run variance of $V_{ij}\left(\pi_{j}\right)-EV_{ij}\left(\pi_{j}\right)$
		and $\beta\left(q\right)\leq e^{-Bq}$ is the $\beta$-mixing
		coefficients. From the lemma 1 in \citet{yoshihara1976limiting} and lemma C.2 in
		\citet{galvao2016smoothed}, we know $\sigma_{q}\left(V_{ij}\left(\pi_{j}\right)-EV_{ij}\left(\pi_{j}\right)\right)\leq C$
		for some finite positive constant $C$ when $\beta\left(q\right)$
		decays exponentially.
		
		We take $q=\left\lfloor n^{\frac{1}{2}}\right\rfloor $ and $s=\frac{n^{\frac{1}{2}-\kappa}}{r_{n}^{\frac{1}{2}}}$,
		so $r=\left\lfloor \frac{n^{\frac{1}{2}}}{2}\right\rfloor $. Then
		we have 
		\begin{align*}
			& P\Biggl\{\sup_{\left\Vert \pi_{j}-\pi_{j}^{0}\right\Vert \leq C_{8}n^{-\kappa}}\left|\frac{1}{n}\sum_{i=1}^{n}V_{ij}\left(\pi_{ij}\right)-EV_{ij}\left(\pi_{j}\right)\right|\\
			& \geq C^{**}\left[\frac{n^{\frac{1}{4}-\frac{\kappa}{2}}}{r_{n}^{\frac{1}{4}}n^{\frac{1}{2}}}+2M_{1}C_{8}\frac{n^{\frac{1}{2}-\kappa}n^{\frac{1}{2}}r_{n}^{\frac{1}{2}}n^{-\kappa}}{r_{n}^{\frac{1}{2}}n}\right]\Biggl\}\\
			\text{\ensuremath{\leq}} & P\Biggl\{\sup_{\left\Vert \pi_{j}-\pi_{j}^{0}\right\Vert \leq C_{8}n^{-\kappa}}\left|\frac{1}{n}\sum_{i=1}^{n}V_{ij}\left(\pi_{ij}\right)-EV_{ij}\left(\pi_{j}\right)\right|\\
			& \geq C^{**}\left[\frac{n^{-\frac{\kappa}{2}}}{n^{\frac{\gamma}{4}}n^{\frac{1}{4}}}+Cn^{-2\kappa}\right]\Biggl\}\\
			\text{\ensuremath{\leq}} & P\Biggl\{\sup_{\left\Vert \pi_{j}-\pi_{j}^{0}\right\Vert \leq C_{8}n^{-\kappa}}\left|\frac{1}{n}\sum_{i=1}^{n}V_{ij}\left(\pi_{ij}\right)-EV_{ij}\left(\pi_{j}\right)\right|\geq C^{***}n^{-2\kappa}\Biggl\}
		\end{align*}
		where the last inequality holds as long as $\frac{\kappa}{2}+\frac{\gamma}{4}+\frac{1}{4}\geq2\kappa\Leftrightarrow\gamma\geq6\kappa-1$,
		which is true in our assumption. Consider $c_{9} = 2C^{***}$,
		we have 
		\begin{align*}
			\Gamma_{4} & =P\Biggl\{\sup_{\left\Vert \pi_{j}-\pi_{j}^{0}\right\Vert \leq C_{8}n^{-\kappa}}\left|\frac{1}{n}\sum_{i=1}^{n}V_{ij}\left(\pi_{ij}\right)-EV_{ij}\left(\pi_{j}\right)\right|\geq\frac{1}{2}c_{9}n^{-2\kappa}\Biggl\}\\
			& \leq c_{12}e^{-c_{13}\left(\frac{n^{1-2\kappa}}{r_{n}}\right)}+c_{14}n^{\frac{1}{2}}e^{-Bn^{\frac{1}{2}}}
		\end{align*}
		
		Hence, 
		\begin{align*}
			P\left(\left\Vert \hat{\pi}_{j}-\pi_{j}^{0}\right\Vert \geq
			C_{8}n^{-\kappa}\right) &  \leq e^{-C_{9}n^{1-4\kappa}}+C_{10}e^{-C_{11}\left(\frac{n^{1-2\kappa}}{r_{n}}\right)}+C_{12}n^{\frac{1}{2}}e^{-Bn^{\frac{1}{2}}}
		\end{align*}
		where $C_9 = c_{11}$, $C_{10} = c_{12}$, $C_{11} = c_{13}$, and $C_{12} = c_{14}$
		
	\end{proof}
	
	\subsection{Proof of lemma \protect\ref{lem:4}}
	
	\begin{proof}
		
		We first rewrite 
		\begin{align*}
			& \frac{1}{n}\sum_{i=1}^{n}\psi_{\tau}\left(Y_{i+1}-X_{i,S_{j}}^{T}\hat{\pi}%
			_{j}\right)\left(X_{i,j}-X_{i,S_{j}}^{T}\hat{\theta}_{j}\right)-E\left[%
			\psi_{\tau}\left(Y_{i+1}-X_{i,S_{j}}^{T}\pi_{j}^{0}\right)X_{i,j}\right] \\
			= & \Delta_{1j}+\Delta_{2j}+\Delta_{3j}
		\end{align*}
		where 
		\begin{align*}
			\Delta_{1j} & =\frac{1}{n}\sum_{i=1}^{n}\psi_{\tau}%
			\left(Y_{i+1}-X_{i,S_{j}}^{T}\pi_{j}^{0}\right)X_{i,j}-E\left[%
			\psi_{\tau}\left(Y-X_{i,S_{j}}^{T}\pi_{j}^{0}\right)X_{i,j}\right] \\
			\Delta_{2j} & =\frac{1}{n}\sum_{i=1}^{n}\left(\psi_{\tau}%
			\left(Y_{i+1}-X_{i,S_{j}}^{T}\hat{\pi}_{j}\right)-\psi_{\tau}%
			\left(Y_{i+1}-X_{i,S_{j}}^{T}\pi_{j}^{0}\right)\right)X_{i,j} \\
			\Delta_{3j} & =-\frac{1}{n}\sum_{i=1}^{n}\psi_{\tau}%
			\left(Y_{i+1}-X_{i,S_{j}}^{T}\hat{\pi}_{j}\right)X_{i,S_{j}}^{T}\hat{\theta}%
			_{j}
		\end{align*}
		
		By assumption \ref{assu:bound}, we have $\left|\psi_{\tau}%
		\left(Y_{i+1}-X_{i,S_{j}}^{T}\pi_{j}^{0}\right)X_{i,j}\right|\leq M_{1}$.
		Since $\psi_{\tau}\left(Y_{i+1}-X_{i,S_{j}}^{T}\pi_{j}^{0}\right)X_{i,j}$ is 
		$\alpha$-mixing, using theorem 1 in \citet{merlevede2009bernstein}, we have 
		
		\begin{align*}
			P\left(\left|\Delta_{1j}\right|\geq c_{15}n^{-\kappa}\right) & \leq e^{-%
				\frac{cc_{15}^{2}n^{2-2\kappa}}{nM_{1}^{2}+M_{1}c_{15}\log n\log\log n}} \\
			& \leq e^{-c_{16}n^{1-2\kappa}}
		\end{align*}
		
		For $\Delta_{2j}$, we first consider there exists $u^{*}$ with $\left\Vert
		u^{*}\right\Vert \leq1$ and some positive constant $c_{17}$, then we use
		Cauchy-Schwarz inequality with assumption \ref{assu:beta-mix} and \ref{assu:density} 
		\begin{align*}
			& \sup_{\left\Vert u\right\Vert
				\leq1}\left|\left(\psi_{\tau}\left(Y_{i+1}-X_{i,S_{j}}^{T}\left(%
			\pi_{j}^{0}+c_{17}n^{-\kappa}u\right)\right)-\psi_{\tau}%
			\left(Y_{i+1}-X_{i,S_{j}}^{T}\pi_{j}^{0}\right)\right)X_{i,j}\right| \\
			= & \sup_{\left\Vert u\right\Vert
				\leq1}\left|\left(\psi_{\tau}\left(Y_{i+1}-X_{i,S_{j}}^{T}\left(%
			\pi_{j}^{0}+c_{17}n^{-\kappa}u^{*}\right)\right)-\psi_{\tau}%
			\left(Y_{i+1}-X_{i,S_{j}}^{T}\pi_{j}^{0}\right)\right)X_{i,j}\right| \\
			\leq & \sup_{\left\Vert u\right\Vert
				\leq1}\left|\int_{X_{i,S_{j}}^{T}\pi_{j}^{0}}^{X_{i,S_{j}}^{T}\left(%
				\pi_{j}^{0}+c_{17}n^{-\kappa}u^{*}\right)}f_{Y_{i+1}|X_{i}}\left(y\right)dy%
			\right|\left|X_{i,j}\right| \\
			\leq & c_{17}n^{-\kappa}\left|X_{i,S_{j}}^{T}u^{*}\right| \\
			\leq & c_{17}M_{1}^{2}r_{n}^{\frac{1}{2}}n^{-\kappa}
		\end{align*}
		
		We define 
		\begin{equation*}
			\Pi_{ij}=\sup_{\left\Vert u\right\Vert
				\leq1}\left|\left(\psi_{\tau}\left(Y_{i+1}-X_{i,S_{j}}^{T}\left(%
			\pi_{j}^{0}+c_{13}n^{-\kappa}u\right)\right)-\psi_{\tau}%
			\left(Y_{i+1}-X_{i,S_{j}}^{T}\pi_{j}^{0}\right)\right)X_{i,j}\right|
		\end{equation*}
		
		Using theorem 1 in \citet{merlevede2009bernstein}, we obtain 
		\begin{align*}
			& P\left(\left|\frac{1}{n}\sum_{i=1}^{n}\Pi_{ij}-E\Pi_{ij}\right|\geq
			c_{18}r_{n}^{\frac{1}{2}}n^{-\kappa}\right) \\
			\leq & e^{-\frac{c\left(c_{18}r_{n}^{\frac{1}{2}}n^{1-\kappa}\right)^{2}}{%
					n\left(c_{17}M_{1}^{2}r_{n}^{\frac{1}{2}}n^{-\kappa}\right)^{2}+%
					\left(c_{17}M_{1}^{2}r_{n}^{\frac{1}{2}}n^{-\kappa}\right)c_{18}r_{n}^{\frac{%
							1}{2}}n^{-\kappa}\log n\log\log n}} \\
			\leq & e^{-c_{19}n}
		\end{align*}
		
		Therefore, since we have $E\Pi_{ij}\leq c_{17}M_{1}^{2}r_{n}^{\frac{1}{2}%
		}n^{-\kappa}$, we show that 
		\begin{align*}
			& P\left(\left|\frac{1}{n}\sum_{i=1}^{n}\Pi_{ij}\right|\geq c_{19}r_{n}^{%
				\frac{1}{2}}n^{-\kappa}\right) \\
			\leq & P\left(\left|\frac{1}{n}\sum_{i=1}^{n}\Pi_{ij}-E\Pi_{ij}\right|+E%
			\Pi_{ij}\geq c_{19}r_{n}^{\frac{1}{2}}n^{-\kappa}\right) \\
			\leq & P\left(\left|\frac{1}{n}\sum_{i=1}^{n}\Pi_{ij}-E\Pi_{ij}\right|\geq
			c_{20}r_{n}^{\frac{1}{2}}n^{-\kappa}\right) \\
			\leq & e^{-c_{21}n}
		\end{align*}
		for some positive constants $c_{19},c_{20},c_{21}$ and the last inequality
		follows the same procedure as above.
		
		Hence, we have for some positive constant $c_{22},c_{23}$ 
		\begin{align*}
			& P\left(\left|\Delta_{2j}\right|\geq c_{22}r_{n}^{\frac{1}{2}%
			}n^{-\kappa}\right) \\
			\leq & P\left(\left|\Delta_{2j}\right|\geq c_{22}r_{n}^{\frac{1}{2}%
			}n^{-\kappa},\left\Vert \hat{\pi}_{j}-\pi_{j}^{0}\right\Vert
			<C_{8}n^{-\kappa}\right) \\
			& +P\left(\left|\Delta_{2j}\right|\geq c_{22}r_{n}^{\frac{1}{2}%
			}n^{-\kappa},\left\Vert \hat{\pi}_{j}-\pi_{j}^{0}\right\Vert \geq
			C_{8}n^{-\kappa}\right) \\
			\leq & P\left(\left|\frac{1}{n}\sum_{i=1}^{n}\Pi_{ij}\right|\geq
			c_{22}r_{n}^{\frac{1}{2}}n^{-\kappa}\right)+P\left(\left\Vert \hat{\pi}%
			_{j}-\pi_{j}^{0}\right\Vert \geq C_{8}n^{-\kappa}\right) \\
			\leq &
			e^{-c_{23}n}+e^{-C_{9}n^{1-4\kappa}}+C_{10}e^{-C_{11}r_{n}^{-1}n^{1-2%
					\kappa}}+C_{12}n^{\frac{1}{2}}e^{-Bn^{\frac{1}{2}}}
		\end{align*}
		where the last two inequalities rely on the result of lemma \ref{lem:3}.
		
		For $\Delta_{3j}$, we first define 
		\begin{equation*}
			g\left(\pi_{j}\right)=\frac{1}{n}\sum_{i=1}^{n}\rho_{\tau}%
			\left(Y_{i+1}-X_{i,S_{j}}^{T}\pi_{j}\right)
		\end{equation*}
		and its subdifferential 
		\begin{equation*}
			\partial g\left(\pi_{j}\right)=\left\{ \partial
			g_{k}\left(\pi_{j}\right):k\in\left\{ 0\right\} \cup S_{j}\right\} ^{T}
		\end{equation*}
		with 
		\begin{equation*}
			\partial g_{k}\left(\pi_{j}\right)=-\frac{1}{n}\sum_{i=1}^{n}\psi_{\tau}%
			\left(Y_{i+1}-X_{i,S_{j}}^{T}\pi_{j}\right)X_{i,k}-\frac{1}{n}%
			\sum_{i=1}^{n}1\left(Y_{i+1}=X_{i,S_{j}}^{T}\pi_{j}\right)v_{i}X_{i,j}
		\end{equation*}
		where $v_{i}\in\left[\tau-1,\tau\right]$.
		
		Since $\hat{\pi}_{j}=\arg\min\frac{1}{n}\sum_{i=1}^{n}\rho_{\tau}%
		\left(Y_{i+1}-X_{i,S_{j}}^{T}\pi_{j}\right)$, we know there exists $%
		v_{i}^{*}\in\left[\tau-1,\tau\right]$ such that $\partial g_{k}\left(\hat{\pi%
		}_{j}\right)=0$. Therefore, we have 
		\begin{align*}
			\Delta_{3j} & =\frac{1}{n}\sum_{i=1}^{n}1\left(Y_{i+1}=X_{i,S_{j}}^{T}\hat{%
				\pi}_{j}\right)v_{i}^{*}X_{i,S_{j}}^{T}\hat{\theta}_{j}
		\end{align*}
		and. by the triangular inequality and assumption \ref{assu:bound}, we know 
		\begin{align*}
			& \left|\frac{1}{n}\sum_{i=1}^{n}1\left(Y_{i+1}=X_{i,S_{j}}^{T}\hat{\pi}%
			_{j}\right)v_{i}^{*}X_{i,S_{j}}^{T}\hat{\theta}_{j}\right| \\
			\leq & \frac{1}{n}\sum_{i=1}^{n}1\left(Y_{i+1}=X_{i,S_{j}}^{T}\hat{\pi}%
			_{j}\right)\left|X_{i,S_{j}}^{T}\hat{\theta}_{j}\right| \\
			\leq & \frac{1}{n}\sum_{i=1}^{n}1\left(Y_{i+1}=X_{i,S_{j}}^{T}\hat{\pi}%
			_{j}\right)\left(\left|X_{i,S_{j}}^{T}\theta_{j}^{0}\right|+%
			\left|X_{i,S_{j}}^{T}\hat{\theta}_{j}-X_{i,S_{j}}^{T}\theta_{j}^{0}\right|%
			\right) \\
			\leq & \frac{1}{n}\sum_{i=1}^{n}1\left(Y_{i+1}=X_{i,S_{j}}^{T}\hat{\pi}%
			_{j}\right)\left(M_{2}+r_{n}^{\frac{1}{2}}M_{1}\left\Vert \hat{\theta}%
			_{j}-\theta_{j}^{0}\right\Vert \right)
		\end{align*}
		
		Using lemma \ref{lem:2} and let $\delta_{n}=\left(\frac{r_{n}}{n}%
		\right)^{-1} $, we know
		
		\begin{equation*}
			P\left(\left\Vert \hat{\theta}_{j}-\theta_{j}^{0}\right\Vert \geq
			C_{6}\right)\leq C_{7}r_{n}e^{-\frac{C_{3}\delta_{n}^{2}}{n}}+e^{-\frac{%
					C_{5}n}{r_{n}^{2}}}=c_{24}r_{n}e^{-c_{25}\frac{n}{r_{n}^{2}}}
		\end{equation*}
		
		Hence, we have 
		\begin{equation*}
			P\left(M_{2}+r_{n}^{\frac{1}{2}}M_{1}\left\Vert \hat{\theta}%
			_{j}-\theta_{j}^{0}\right\Vert \geq M_{2}+C_{6}M_{1}r_{n}^{\frac{1}{2}%
			}\right)\leq c_{24}r_{n}e^{-c_{25}\frac{n}{r_{n}^{2}}}
		\end{equation*}
		
		Since $P\left(Y_{i+1}=X_{i,S_{j}}^{T}\hat{\pi}_{j}\right)=0$ and $P\left(%
		\frac{1}{n}\sum_{i=1}^{n}1\left(Y_{i+1}=X_{i,S_{j}}^{T}\hat{\pi}%
		_{j}\right)>\varepsilon\right)=0$ for any $\varepsilon>0$. We let $%
		\varepsilon=\frac{1}{r_{n}^{\frac{1}{2}}n}$, then we have 
		\begin{align*}
			& P\left(\left|\Delta_{3j}\right|\geq\frac{1}{r_{n}^{\frac{1}{2}}n}%
			\left(M_{2}+C_{6}M_{1}r_{n}^{\frac{1}{2}}\right)\right) \\
			\leq & P\left(\left|\frac{1}{n}\sum_{i=1}^{n}1\left(Y_{i+1}=X_{i,S_{j}}^{T}%
			\hat{\pi}_{j}\right)\right|\left|\left(M_{2}+r_{n}^{\frac{1}{2}%
			}M_{1}\left\Vert \hat{\theta}_{j}-\theta_{j}^{0}\right\Vert
			\right)\right|\geq\frac{1}{r_{n}^{\frac{1}{2}}n}%
			\left(M_{2}+C_{6}M_{1}r_{n}^{\frac{1}{2}}\right)\right) \\
			\leq & c_{24}r_{n}e^{-c_{25}\frac{n}{r_{n}^{2}}}
		\end{align*}
		
		Finally, we combine the probabilistic bounds for $\Delta_{1j}$, $\Delta_{2j}$
		and $\Delta_{3j}$, so we obtain 
		\begin{align*}
			& P\left\{ \left|\frac{1}{n}\sum_{i=1}^{n}\psi_{\tau}%
			\left(Y_{i+1}-X_{i,S_{j}}^{T}\hat{\pi}_{j}\right)%
			\left(X_{i,j}-X_{i,S_{j}}^{T}\hat{\theta}_{j}\right)-E\left[%
			\psi_{\tau}\left(Y_{i+1}-X_{i,S_{j}}^{T}\pi_{j}^{0}\right)X_{i,j}\right]%
			\right|\geq C_{13}r_{n}^{\frac{1}{2}}n^{-\kappa}\right\} \\
			\leq & P\{\left|\frac{1}{n}\sum_{i=1}^{n}\psi_{\tau}%
			\left(Y_{i+1}-X_{i,S_{j}}^{T}\hat{\pi}_{j}\right)%
			\left(X_{i,j}-X_{i,S_{j}}^{T}\hat{\theta}_{j}\right)-E\left[%
			\psi_{\tau}\left(Y_{i+1}-X_{i,S_{j}}^{T}\pi_{j}^{0}\right)X_{i,j}\right]%
			\right| \\
			& \geq c_{15}n^{-\kappa}+c_{22}r_{n}^{\frac{1}{2}}n^{-\kappa}+\frac{1}{%
				r_{n}^{\frac{1}{2}}n}\left(M_{2}+C_{6}M_{1}r_{n}^{\frac{1}{2}}\right)\} \\
			\leq &
			e^{-c_{16}n^{1-2\kappa}}+e^{-c_{23}n}+e^{-C_{9}n^{1-4%
					\kappa}}+C_{10}e^{-C_{11}r_{n}^{-1}n^{1-2\kappa}}+C_{12}n^{\frac{1}{2}%
			}e^{-Bn^{\frac{1}{2}}}+c_{24}r_{n}e^{-c_{25}\frac{n}{r_{n}^{2}}} \\
			\leq &
			C_{14}e^{-C_{15}n^{1-4\kappa}}+C_{10}e^{-C_{11}r_{n}^{-1}n^{1-2\kappa}}+C_{12}n^{%
				\frac{1}{2}}e^{-Bn^{\frac{1}{2}}}+C_{16}r_{n}e^{-C_{17}\frac{n}{%
					r_{n}^{2}}}
		\end{align*}
		where $C_{16}=c_{22}$ and $C_{17}=c_{23}$.
		
	\end{proof}
	
	\subsection{Proof of lemma \protect\ref{lem:5}}
	
	\begin{proof}
		
		Since $\sigma_{j}^{2}=var\left(X_{i,j}-X_{i,S_{j}}^{T}\theta_{j}^{0}\right)$
		and $\hat{\sigma}_{j}^{2}=\frac{1}{n}\sum_{i=1}^{n}%
		\left(X_{i,j}-X_{i,S_{j}}^{T}\hat{\theta}_{j}\right)^{2}$, we can rewrite 
		\begin{equation*}
			\left|\hat{\sigma}_{j}^{2}-\sigma_{j}^{2}\right|\leq\Gamma_{5j}+\Gamma_{6j}%
			\left(\hat{\theta}_{j}\right)
		\end{equation*}
		where 
		\begin{align*}
			\Gamma_{5j} & =\left|\frac{1}{n}\sum_{i=1}^{n}\left(X_{i,j}-X_{i,S_{j}}^{T}%
			\theta_{j}^{0}\right)^{2}-E\left(X_{i,j}-X_{i,S_{j}}^{T}\theta_{j}^{0}%
			\right)^{2}\right| \\
			\Gamma_{6j}\left(\hat{\theta}_{j}\right) & =\left|\frac{1}{n}%
			\sum_{i=1}^{n}\left(X_{i,j}-X_{i,S_{j}}^{T}\hat{\theta}_{j}\right)^{2}-\frac{%
				1}{n}\sum_{i=1}^{n}\left(X_{i,j}-X_{i,S_{j}}^{T}\theta_{j}^{0}\right)^{2}%
			\right|
		\end{align*}
		
		For $\Gamma_{5j}$, we use theorem 1 in \citet{merlevede2009bernstein} and we
		know 
		\begin{equation*}
			\left|\left(X_{i,j}-X_{i,S_{j}}^{T}\theta_{j}^{0}\right)^{2}-E%
			\left(X_{i,j}-X_{i,S_{j}}^{T}\theta_{j}^{0}\right)^{2}\right|\leq M
		\end{equation*}
		for some universal constant $M$ by assumption \ref{assu:bound}. So we have 
		\begin{align*}
			P\left(\Gamma_{5j}\geq c_{26}r_{n}^{\frac{1}{2}}n^{-\kappa}\right) & \leq
			e^{-\frac{c\left(nc_{26}r_{n}^{\frac{1}{2}}n^{-\kappa}\right)^{2}}{%
					-nM+Mnc_{26}r_{n}^{\frac{1}{2}}n^{-\kappa}\log n\log\log n}} \\
			& \leq e^{-c_{27}r_{n}n^{1-2\kappa}}
		\end{align*}

		For $\Gamma_{6j}\left(\hat{\theta}_{j}\right)$, we can rewrite it as 
		\begin{align*}
			\Gamma_{6j}\left(\hat{\theta}_{j}\right) & =\left|\frac{1}{n}%
			\sum_{i=1}^{n}\left(\left(X_{i,j}-X_{i,S_{j}}^{T}\hat{\theta}%
			_{j}\right)+\left(X_{i,j}-X_{i,S_{j}}^{T}\theta_{j}^{0}\right)\right)%
			\left(X_{i,S_{j}}^{T}\left(\hat{\theta}_{j}-\theta_{j}^{0}\right)\right)%
			\right| \\
			& =\left|\frac{1}{n}\sum_{i=1}^{n}\left(2\left(X_{i,j}-X_{i,S_{j}}^{T}%
			\theta_{j}^{0}\right)+X_{i,S_{j}}^{T}\left(\hat{\theta}_{j}-\theta_{j}^{0}%
			\right)\right)\left(X_{i,S_{j}}^{T}\left(\hat{\theta}_{j}-\theta_{j}^{0}%
			\right)\right)\right| \\
			& \leq\left|\frac{1}{n}\sum_{i=1}^{n}\left(2\left(X_{i,j}-X_{i,S_{j}}^{T}%
			\theta_{j}^{0}\right)X_{i,S_{j}}^{T}\left(\hat{\theta}_{j}-\theta_{j}^{0}%
			\right)\right)\right|+\left(\hat{\theta}_{j}-\theta_{j}^{0}\right)^{T}\left(%
			\frac{1}{n}\sum_{i=1}^{n}X_{i,S_{j}}X_{i,S_{j}}^{T}\right)\left(\hat{\theta}%
			_{j}-\theta_{j}^{0}\right) \\
			& :=\Gamma_{7j}+\Gamma_{8j}\left(\hat{\theta}_{j}\right)
		\end{align*}
		
		For $\Gamma_{8j}\left(\hat{\theta}_{j}\right)$, from the proof of lemma \ref%
		{lem:2}, we know
		
		\begin{equation*}
			P\left(\left\Vert \frac{1}{n}\sum_{i=1}^{n}X_{i,S_{j}}X_{i,S_{j}}^{T}-E%
			\left(X_{i,S_{j}}X_{i,S_{j}}^{T}\right)\right\Vert \geq r_{n}\frac{%
				c_{3}\delta_{1}}{n}\right)\leq e^{-\frac{C_{4}\delta_{1}^{^{2}}}{n}}
		\end{equation*}
		
		Together with the inequality $\left|\lambda_{\max}\left(A\right)-\lambda_{%
			\max}\left(B\right)\right|\leq\left\Vert A-B\right\Vert $ for symmetric
		matrices $A$ and $B$, we obtain
		
		\begin{equation*}
			P\left(\left\Vert \lambda_{\max}\left(\frac{1}{n}%
			\sum_{i=1}^{n}X_{i,S_{j}}X_{i,S_{j}}^{T}\right)-\lambda_{\max}\left(E%
			\left(X_{i,S_{j}}X_{i,S_{j}}^{T}\right)\right)\right\Vert \geq r_{n}\frac{%
				c_{1}\delta_{1}}{n}\right)\leq e^{-\frac{c_{4}\delta_{1}^{2}}{n}}
		\end{equation*}
		
		Similarly, we let $\delta_{1}=\frac{c_{28}n}{c_{3}r_{n}}m\leq\frac{c_{28}n}{%
			c_{3}r_{n}}\lambda_{\max}\left(E\left(X_{i,S_{j}}X_{i,S_{j}}^{T}\right)%
		\right)$ for some constant $c_{28}$ and denote $c_{29}=\frac{c_{28}^{2}}{%
			c_{3}^{2}c_{4}^{\frac{1}{2}}}$, then we obtain 
		\begin{equation*}
			P\left(\left|\lambda_{\max}\left(\frac{1}{n}%
			\sum_{i=1}^{n}X_{i,S_{j}}X_{i,S_{j}}^{T}\right)-\lambda_{\max}\left(E%
			\left(X_{i,S_{j}}X_{i,S_{j}}^{T}\right)\right)\right|\geq
			c_{28}\lambda_{\max}\left(E\left(X_{i,S_{j}}X_{i,S_{j}}^{T}\right)\right)%
			\right)\leq e^{-\frac{c_{29}n}{r_{n}^{2}}}
		\end{equation*}
		which indicates 
		\begin{equation*}
			P\left(\lambda_{\max}\left(\frac{1}{n}%
			\sum_{i=1}^{n}X_{i,S_{j}}X_{i,S_{j}}^{T}\right)\geq\left(1+c_{28}\right)M%
			\right)\leq e^{-\frac{c_{29}n}{r_{n}^{2}}}
		\end{equation*}
		by assumption \ref{assu:eigen}.
		
		From lemma \ref{lem:2}, we know
		
		\begin{equation*}
			P\left(\left\Vert \hat{\theta}_{j}-\theta_{j}^{0}\right\Vert \geq C_{6}\frac{%
				r_{n}\delta_{n}}{n}\right)\leq C_{7}r_{n}e^{-\frac{C_{3}\delta_{n}^{2}}{n}%
			}+e^{-\frac{C_{5}n}{r_{n}^{2}}}
		\end{equation*}
		
		We let $\delta_{n}=c_{30}r_{n}^{-\frac{1}{2}}n^{1-\kappa}$ for some positive
		constant $c_{30}$ and we obtain 
		\begin{equation*}
			P\left(\left\Vert \hat{\theta}_{j}-\theta_{j}^{0}\right\Vert \geq
			C_{6}c_{30}r_{n}^{\frac{1}{2}}n^{-\kappa}\right)\leq C_{7}r_{n}e^{-\frac{%
					C_{3}c_{30}^{2}n^{1-2\kappa}}{r_{n}}}+e^{-\frac{C_{5}n}{r_{n}^{2}}}
		\end{equation*}
		
		Therefore, we can show 
		\begin{align*}
			& P\left(\Gamma_{8j}\left(\hat{\theta}_{j}\right)\geq\left(1+c_{28}\right)M%
			\left(C_{6}c_{30}r_{n}^{\frac{1}{2}}n^{-\kappa}\right)^{2}\right) \\
			\leq & e^{-\frac{c_{29}n}{r_{n}^{2}}}+C_{7}r_{n}e^{-\frac{%
					C_{3}c_{23}^{2}n^{1-2\kappa}}{r_{n}}}+e^{-\frac{C_{5}n}{r_{n}^{2}}}
		\end{align*}
		
		As for $\Gamma_{7j}$, let $\theta_{j}=\theta_{j}^{0}+c_{31}r_{n}^{\frac{1}{2}%
		}n^{-\kappa}u$ where $c_{31}=C_{6}c_{26}$, $u\in\mathbb{R}%
		^{\left|S_{j}\right|}$ and $\left\Vert u\right\Vert \leq1$. We then define 
		\begin{equation*}
			\Phi_{j}\left(u\right)=\frac{1}{n}\sum_{i=1}^{n}%
			\left(X_{i,j}-X_{i,S_{j}}^{T}\theta_{j}^{0}\right)X_{i,S_{j}}^{T}\left(\hat{%
				\theta}_{j}-\theta_{j}^{0}\right)
		\end{equation*}
		
		From assumption \ref{assu:bound}, we know $\left|\left(X_{i,j}-X_{i,S_{j}}^{T}%
		\theta_{j}^{0}\right)X_{i,S_{j}}^{T}\left(\hat{\theta}_{j}-\theta_{j}^{0}%
		\right)\right|\leq Mr_{n}n^{-\kappa}$ for some universal constant $M$. Using
		theorem 1 in \citet{merlevede2009bernstein}, we can obtain 
		\begin{align*}
			P\left(\left|\Phi_{j}\left(u\right)\right|\geq c_{32}r_{n}^{\frac{1}{2}%
			}n^{-\kappa}\right) & \leq e^{-\frac{c\left(c_{32}^{2}nr_{n}^{\frac{1}{2}%
					}n^{-\kappa}\right)^{2}}{n\left(Mr_{n}n^{-\kappa}\right)^{2}+Mr_{n}n^{-%
						\kappa}c_{32}nr_{n}^{\frac{1}{2}}n^{-\kappa}\log n\log\log n}} \\
			& \leq e^{-\frac{c_{33}n}{r_{n}}}
		\end{align*}
		
		Then we partition $\Lambda=\left\{ u:u\in\mathbb{R}^{\left|S_{j}\right|},%
		\left\Vert u\right\Vert \leq1\right\} $ as a union of $l_{n}$ disjoint
		subsets $\Lambda_{1},\dots,\Lambda_{l_{n}}$. Each subset has equal spaces in
		each direction of $u$. Therefore, we know $\sup_{u,u^{^{\prime
			}}\in\Lambda_{k}}\left\Vert u-u^{^{\prime }}\right\Vert \leq\frac{\sqrt{r_{n}%
		}}{l_{n}^{\frac{1}{\left|S_{j}\right|}}}$ for all $k\in\left\{
		1,\dots,l_{n}\right\} $. Hence, for $u_{k}\in\Lambda_{k}$, we have 
		\begin{equation*}
			\sup_{u\in\Lambda}\left|\Phi_{j}\left(u\right)\right|\leq\sup_{k}\left|%
			\Phi_{j}\left(u_{k}\right)\right|+\sup_{k}\sup_{u\in\Lambda_{k}}\left|%
			\Phi_{j}\left(u\right)-\Phi_{j}\left(u_{k}\right)\right|
		\end{equation*}
		
		By the previous inequality and the Bonferroni bound, we show that 
		\begin{equation*}
			P\left(\sup_{k}\left|\Phi_{j}\left(u_{k}\right)\right|\geq c_{32}r_{n}^{%
				\frac{1}{2}}n^{-\kappa}\right)\leq l_{n}e^{-\frac{c_{33}n}{r_{n}}}
		\end{equation*}
		
		Moreover, we know 
		\begin{align*}
			&
			\sup_{k}\sup_{u\in\Lambda_{k}}\left|\Phi_{j}\left(u\right)-\Phi_{j}%
			\left(u_{k}\right)\right| \\
			= & \sup_{k}\sup_{u\in\Lambda_{k}}\left|\frac{1}{n}\sum_{i=1}^{n}%
			\left(X_{i,j}-X_{i,S_{j}}^{T}\theta_{j}^{0}\right)X_{i,S_{j}}^{T}%
			\left(c_{31}r_{n}^{\frac{1}{2}}n^{-\kappa}\left(u-u_{k}\right)\right)\right|
			\\
			\leq &
			\left(M_{1}+M_{2}\right)M_{1}c_{31}r_{n}n^{-\kappa}\sup_{k}\sup_{u\in%
				\Lambda_{k}}\left\Vert u-u_{k}\right\Vert \\
			\leq & c_{34}r_{n}^{\frac{3}{2}}n^{-\kappa}\frac{1}{l_{n}^{\frac{1}{%
						\left|S_{j}\right|}}}
		\end{align*}
		where $c_{30}=\left(M_{1}+M_{2}\right)M_{1}c_{31}$.
		
		Letting $l_{n}^{\frac{1}{\left|S_{j}\right|}}=r_{n}$, we have 
		\begin{equation*}
			\sup_{k}\sup_{u\in\Lambda_{k}}\left|\Phi_{j}\left(u\right)-\Phi_{j}%
			\left(u_{k}\right)\right|\leq c_{34}r_{n}^{\frac{1}{2}}n^{-\kappa}
		\end{equation*}
		
		Hence, denoting $c_{35}=c_{32}+c_{34}$, we obtain 
		\begin{equation*}
			P\left(\sup_{u\in\Lambda}\left|\Phi_{j}\left(u\right)\right|\geq
			c_{35}r_{n}^{\frac{1}{2}}n^{-\kappa}\right)\leq r_{n}^{r_{n}}e^{-\frac{%
					c_{33}n}{r_{n}}}\leq e^{-\frac{c_{33}n}{r_{n}}+r_{n}\log r_{n}}\leq e^{-%
				\frac{c_{36}n}{r_{n}}}
		\end{equation*}
		which indicates 
		\begin{equation*}
			P\left(\left|\Gamma_{7j}\right|\geq c_{37}r_{n}^{\frac{1}{2}%
			}n^{-\kappa}\right)\leq e^{-\frac{c_{36}n}{r_{n}}}
		\end{equation*}
		for some positive constant $c_{37}$ under the event $\left\Vert \hat{\theta}%
		_{j}-\theta_{j}^{0}\right\Vert \leq c_{31}r_{n}^{\frac{1}{2}}n^{-\kappa}$.
		
		Hence, by the probabilistic bound of $\left\Vert \hat{\theta}%
		_{j}-\theta_{j}^{0}\right\Vert $ , we have 
		\begin{equation*}
			P\left(\left|\Gamma_{7j}\right|\geq c_{37}r_{n}^{\frac{1}{2}%
			}n^{-\kappa}\right)\leq e^{-\frac{c_{36}n}{r_{n}}}+C_{7}r_{n}e^{-\frac{%
					C_{3}c_{30}^{2}n^{1-2\kappa}}{r_{n}}}+e^{-\frac{C_{5}n}{r_{n}^{2}}}
		\end{equation*}
		
		By the assumption that $r_{n}^{\frac{1}{2}}n^{-\kappa}=o\left(1\right)$,
		combining the results of $\Gamma_{7j}$ and $\Gamma_{8j}\left(\hat{\theta}%
		_{j}\right)$, we obtain
		
		\begin{align*}
			& P\left(\left|\Gamma_{6j}\left(\hat{\theta}_{j}\right)\right|\geq
			c_{38}r_{n}^{\frac{1}{2}}n^{-\kappa}\right) \\
			\leq & e^{-\frac{c_{36}n}{r_{n}}}+C_{7}r_{n}e^{-\frac{C_{3}c_{30}^{2}n^{1-2%
						\kappa}}{r_{n}}}+e^{-\frac{C_{5}n}{r_{n}^{2}}}+e^{-\frac{c_{29}n}{r_{n}^{2}}%
			}+C_{7}r_{n}e^{-\frac{C_{3}c_{30}^{2}n^{1-2\kappa}}{r_{n}}}+e^{-\frac{C_{5}n%
				}{r_{n}^{2}}} \\
			\leq & e^{-\frac{c_{36}n}{r_{n}}}+c_{39}e^{-\frac{c_{40}n}{r_{n}^{2}}%
			}+c_{41}r_{n}e^{-\frac{c_{42}n^{1-2\kappa}}{r_{n}}}
		\end{align*}
		
		Therefore, we have
		
		\begin{equation*}
			P\left(\left|\hat{\sigma}_{j}^{2}-\sigma_{j}^{2}\right|\geq C_{18}r_{n}^{%
				\frac{1}{2}}n^{-\kappa}\right)\leq e^{-\frac{C_{19}n}{r_{n}}}+C_{20}e^{-%
				\frac{C_{21}n}{r_{n}^{2}}}+C_{22}r_{n}e^{-\frac{C_{23}n^{1-2\kappa}}{r_{n}}}
		\end{equation*}
		where $C_{19}=c_{36}$, $C_{20}=c_{39}$, $C_{21}=c_{40}$, $C_{22}=c_{41}$, and $C_{23}=c_{42}$.
		
		In the end, by assumption \ref{assu:rn}, we know $r_{n}^{\frac{1}{2}}n^{-\kappa}=o%
		\left(1\right)$ so $C_{18}r_{n}^{\frac{1}{2}}n^{-\kappa}\leq a\sigma_{j}^{2}$
		for some positive number $a\in\left(0,1\right)$, hence we have 
		\begin{equation*}
			P\left(\left|\hat{\sigma}_{j}^{2}-\sigma_{j}^{2}\right|\geq
			a\sigma_{j}^{2}\right)\leq e^{-\frac{C_{19}n}{r_{n}}}+C_{20}e^{-\frac{C_{21}n%
				}{r_{n}^{2}}}+C_{22}r_{n}e^{-\frac{C_{23}n^{1-2\kappa}}{r_{n}}}
		\end{equation*}
	\end{proof}

	\subsection{Proof of theorem \ref{thm:1}}
	
	\begin{proof}
		We know 
		\begin{align*}
			\left|\widehat{qpcor_{\tau}}\left(Y_{i+1},X_{i,j}|X_{i,S_{j}}\right)-qpcor_{%
				\tau}\left(Y_{i+1},X_{i,j}|X_{i,S_{j}}\right)\right| & =\left|\left(\hat{%
				\sigma}_{j}^{2}\right)^{-1}\left(\phi_{jn}-\phi_{j}\right)-\left(\hat{\sigma}%
			_{j}^{2}\right)^{-1}\left(\sigma_{j}^{2}\right)^{-1}\phi_{j}\left(\hat{\sigma%
			}_{j}^{2}-\sigma_{j}^{2}\right)\right| \\
			& \leq\left(\hat{\sigma}_{j}^{2}\right)^{-1}\left|\phi_{jn}-\phi_{j}\right|+%
			\left(\hat{\sigma}_{j}^{2}\right)^{-1}\left(\sigma_{j}^{2}\right)^{-1}\left|%
			\phi_{j}\right|\left|\hat{\sigma}_{j}^{2}-\sigma_{j}^{2}\right|
		\end{align*}
		where $\phi_{jn}=\frac{1}{n}\sum_{i=1}^{n}\psi_{\tau}%
		\left(Y_{i+1}-X_{i,S_{j}}^{T}\hat{\pi}_{j}\right)%
		\left(X_{i,j}-X_{i,S_{j}}^{T}\hat{\theta}_{j}\right)$ and $%
		\phi_{j}=E\left(\psi_{\tau}\left(Y_{i+1}-X_{i,S_{j}}^{T}\pi_{j}^{0}%
		\right)X_{i,j}\right)$
		
		Using the fact that $\left|a^{-1}-b^{-1}\right|\geq
		k^{*}b^{-1}\implies\left|a-b\right|\geq kb$ with $k^{*}=\frac{1}{1-k}-1$ for
		any $k\in\left(0,1\right)$, we can show for some constant $%
		c_{\sigma}^{-1}\geq\left(\sigma_{j}^{2}\right)^{-1}$ 
		\begin{align*}
			P\left(\left(\hat{\sigma}_{j}^{2}\right)^{-1}\geq\left(1+k^{*}\right)c_{%
				\sigma}^{-1}\right) & \leq P\left(\left(\hat{\sigma}_{j}^{2}\right)^{-1}\geq%
			\left(1+k^{*}\right)\left(\sigma_{j}^{2}\right)^{-1}\right) \\
			& \leq P\left(\left|\left(\hat{\sigma}_{j}^{2}\right)^{-1}-\left(%
			\sigma_{j}^{2}\right)\right|\geq k^{*}\left(\sigma_{j}^{2}\right)^{-1}\right)
			\\
			& \leq P\left(\left|\hat{\sigma}_{j}^{2}-\sigma_{j}^{2}\right|\geq
			k\sigma_{j}^{2}\right) \\
			& \leq e^{-\frac{C_{19}n}{r_{n}}}+C_{20}e^{-\frac{C_{21}n}{r_{n}^{2}}%
			}+C_{22}r_{n}e^{-\frac{C_{23}n^{1-2\kappa}}{r_{n}}}
		\end{align*}
		where the last inequality is based on lemma \ref{lem:5}.
		
		From lemma \ref{lem:4}, we know
		
		\begin{align*}
			P\left(\left|\phi_{jn}-\phi_{j}\right|\geq C_{13}r_{n}^{\frac{1}{2}%
			}n^{-\kappa}\right) & \leq
			C_{14}e^{-C_{15}n^{1-4\kappa}}+C_{10}e^{-C_{11}r_{n}^{-1}n^{1-2\kappa}}+C_{12}n^{%
				\frac{1}{2}}e^{-Bn^{\frac{1}{2}}}+C_{16}r_{n}e^{-C_{17}\frac{n}{%
					r_{n}^{2}}}
		\end{align*}
		
		Together with the previous inequality, we show 
		\begin{align*}
			& P\left(\left(\hat{\sigma}_{j}^{2}\right)^{-1}\left|\phi_{jn}-\phi_{j}%
			\right|\geq\left(1+k^{*}\right)c_{\sigma}^{-1}C_{13}r_{n}^{\frac{1}{2}%
			}n^{-\kappa}\right) \\
			\leq & e^{-\frac{C_{19}n}{r_{n}}}+C_{20}e^{-\frac{C_{21}n}{r_{n}^{2}}%
			}+C_{22}r_{n}e^{-\frac{C_{23}n^{1-2\kappa}}{r_{n}}}+C_{14}e^{-C_{15}n^{1-4%
					\kappa}}+C_{10}e^{-C_{11}r_{n}^{-1}n^{1-2\kappa}}+C_{12}n^{\frac{1}{2}%
			}e^{-Bn^{\frac{1}{2}}}+C_{16}r_{n}e^{-C_{17}\frac{n}{r_{n}^{2}}} \\
			\leq &
			C_{14}e^{-C_{5}n^{1-4\kappa}}+c_{42}r_{n}e^{-c_{43}r_{n}^{-1}n^{1-2%
					\kappa}}+C_{12}n^{\frac{1}{2}}e^{-Bn^{\frac{1}{2}%
			}}+c_{45}r_{n}e^{-c_{46}\frac{n}{r_{n}^{2}}}
		\end{align*}
		
		Moreover, using lemma \ref{lem:5} with $\left|\phi_{j}\right|\leq M_{1}$ and 
		$\left(\sigma_{j}^{2}\right)^{-1}\leq c_{\sigma}^{-1}$, 
		\begin{align*}
			& P\left(\left(\hat{\sigma}_{j}^{2}\right)^{-1}\left(\sigma_{j}^{2}%
			\right)^{-1}\left|\phi_{j}\right|\left|\hat{\sigma}_{j}^{2}-\sigma_{j}^{2}%
			\right|\geq\left(1+k^{*}\right)c_{\sigma}^{-1}c_{%
				\sigma}^{-1}M_{1}C_{18}r_{n}^{\frac{1}{2}}n^{-\kappa}\right) \\
			\leq & e^{-\frac{C_{19}n}{r_{n}}}+C_{20}e^{-\frac{C_{21}n}{r_{n}^{2}}%
			}+C_{22}r_{n}e^{-\frac{C_{23}n^{1-2\kappa}}{r_{n}}}
		\end{align*}
		
		Hence, by combining two probabilistic bounds above and rearranging those
		positive constants into $%
		C_{i}^{*}$s and $%
		\tilde{C}_{i}$s , we obtain 
		\begin{align*}
			& P\left(\left|\widehat{qpcor_{\tau}}\left(Y_{i+1},X_{i,j}|X_{i,S_{j}}%
			\right)-qpcor_{\tau}\left(Y_{i+1},X_{i,j}|X_{i,S_{j}}\right)\right|\geq
			C_{1}^{*}r_{n}^{\frac{1}{2}}n^{-\kappa}\right) \\
			\leq & e^{-\frac{C_{2}^{*}n}{r_{n}}}+C_{3}^{*}r_{n}e^{-C_{4}^{*}\frac{n}{%
					r_{n}^{2}}}+C_{5}^{*}r_{n}e^{-\frac{C_{6}^{*}n^{1-2\kappa}}{r_{n}}%
			}+C_{7}^{*}e^{-C_{8}^{*}n^{1-4\kappa}}+C_{9}^{*}n^{\frac{1}{2}%
			}e^{-Bn^{\frac{1}{2}}} \\
			\leq &
			\tilde{C}_{1}r_{n}e^{-\frac{\tilde{C}_{2}n^{1-2\kappa}}{r_{n}}%
			}+\tilde{C}_{3}e^{-\tilde{C}_{4} n^{1-4\kappa}}+\tilde{C}_{5} n^{\frac{1}{2}%
			}e^{-Bn^{\frac{1}{2}}}
		\end{align*}
		
		Therefore,
		
		\begin{align*}
			& P\left(\sup_{1\leq j\leq p_{n}}\left|\widehat{qpcor_{\tau}}%
			\left(Y_{i+1},X_{i,j}|X_{i,S_{j}}\right)-qpcor_{\tau}%
			\left(Y_{i+1},X_{i,j}|X_{i,S_{j}}\right)\right|\geq C_{1}^{*}r_{n}^{\frac{1}{%
					2}}n^{-\kappa}\right) \\
			\leq & p_{n}\left(\tilde{C}_{1}r_{n}e^{-\frac{\tilde{C}_{2}n^{1-2\kappa}}{r_{n}}%
			}+\tilde{C}_{3}e^{-\tilde{C}_{4} n^{1-4\kappa}}+\tilde{C}_{5} n^{\frac{1}{2}%
			}e^{-Bn^{\frac{1}{2}}}\right)
		\end{align*}
		
	\end{proof}
	
	\subsection{Proof of theorem \ref{thm:2}}
	
	\begin{proof}
		
		If we consider the set 
		\begin{equation*}
			E_{n}=\left\{ \sup_{j\in M_{*}}\left|\widehat{qpcor_{\tau}}%
			\left(Y_{i+1},X_{i,j}|X_{i,S_{j}}\right)-qpcor_{\tau}%
			\left(Y_{i+1},X_{i,j}|X_{i,S_{j}}\right)\right|\leq\frac{1}{2}%
			C_{1}^{*}r_{n}^{\frac{1}{2}}n^{-\kappa}\right\}
		\end{equation*}
		with the assumption \ref{assu:identify} that $\min_{j\in
			M_{*}}\left|qpcor_{\tau}\left(Y_{i+1},X_{i,j}|X_{i,S_{j}}\right)\right|\geq%
		\frac{1}{2}C_{1}^{*}r_{n}^{\frac{1}{2}}n^{-\kappa}$ for $C_1^{*} = 2C_0$. We have $\left|\widehat{%
			qpcor_{\tau}}\left(Y_{i+1},X_{i,j}|X_{i,S_{j}}\right)\right|\geq
		C_{1}^{*}r_{n}^{\frac{1}{2}}n^{-\kappa}$. Since $v_{n}=C_{1}^{*}r_{n}^{\frac{%
				1}{2}}n^{-\kappa}$, by the definition of $\hat{M}_{v_{n}}$, we know $%
		M_{*}\subset\hat{M}_{v_{n}}$ on the set $E_{n}$. Therefore, after knowing
		that $s_{n}$ is the number of nonzero coefficients, we have 
		\begin{align*}
			P\left(M_{*}\subset\hat{M}_{v_{n}}\right) & \geq1-P\left(E_{n}^{C}\right) \\
			& \geq1-s_{n}\left(\tilde{C}_{1}r_{n}e^{-\frac{\tilde{C}_{2}n^{1-2\kappa}}{r_{n}}%
			}+\tilde{C}_{3}e^{-\tilde{C}_{4} n^{1-4\kappa}}+\tilde{C}_{5} n^{\frac{1}{2}%
			}e^{-Bn^{\frac{1}{2}}}\right)
		\end{align*}
		by the Bonferroni bound and theorem 1.
		
	\end{proof}

	\subsection{Proof of proposition \ref{prop:1}}
	\begin{proof}
		By the proof of theorem \ref{thm:2} and our assumption \ref{assu:sel-cons} to control the sample QPC of variables with nonzero quantile regression coefficients, the result is straightforward.
	\end{proof}
	
	\subsection{Proof of proposition \ref{prop:2}}
	\begin{proof}
		Since $\sum_{j=1}^{p} \left|qpcor_{\tau}\left(
		Y_{i+1},X_{i,j} | X_{i,S_j}
		\right)  \}\right| = O\left(n^{\zeta}\right)$ in assumption \ref{assu:qpc_control}, we know $\left|\{ j:qpcor\left(Y_{i+1},X_{i,j}|X_{i,S_j}\right) \geq C_1^{*} r_n^{\frac{1}{2}} n^{-\kappa} \}\right| \leq O\left( r_n^{\frac{1}{2}} n^{\zeta + \kappa}  \right) =  O\left( n^{\zeta+\kappa-\frac{\gamma}{2}} \right)$.
		Consider the event set 
		\[
		\Omega_{n}=\left\{ \sup_{1\leq j\leq p}\left|\widehat{qpcor}_{\tau}\left(Y_{i+1},X_{i,j}|X_{i,S_{j}}\right)-qpcor_{\tau}\left(Y_{i+1},X_{i,j}|X_{i,S_{j}}\right)\right|\right\} \leq C_{1}^{*}r_{n}^{\frac{1}{2}}n^{-\kappa}
		\]
		, we must have 
		\begin{align*}
			& \left|\left\{ j:\left|\widehat{qpcor}_{\tau}\left(Y_{i+1},X_{i,j}|X_{i,S_{j}}\right)\right|\geq2C_{1}^{*}r_{n}^{\frac{1}{2}}n^{-\kappa}\right\} \right|\\
			\leq & \left|\left\{ j:\left|qpcor_{\tau}\left(Y_{i+1},X_{i,j}|X_{i,S_{j}}\right)\right|\geq C_{1}^{*}r_{n}^{\frac{1}{2}}n^{-\kappa}\right\} \right|\\
			\leq & C_{1}^{**}n^{\zeta+\kappa-\frac{\gamma}{2}}
		\end{align*}
		and $\left|\hat{M}_{\nu_{n}}\right|\leq C_{1}^{**}n^{\zeta+\kappa-\frac{\gamma}{2}}$ with $\nu_n = C_0r_{n}^{\frac{1}{2}}n^{-\kappa}$ and $C_0 \geq 2 C_1^{*}$.
		By our theorem \ref{thm:1}, we have 
		\[
		P\left(\Omega_{n}\right)\geq1-p_{n}A_{n}
		\]
		Hence, 
		\[
		P\left(\left|\hat{M}_{\nu_{n}}\right|\leq C_{1}^{**}n^{\zeta+\kappa-\frac{\gamma}{2}}\right)\geq1-p_{n}A_{n}
		\]
		
	\end{proof}

\end{document}